\documentclass[fleqn,usenatbib]{mnras}

\usepackage{newtxtext,newtxmath}
\usepackage[T1]{fontenc}

\DeclareRobustCommand{\VAN}[3]{#2}
\let\VANthebibliography\thebibliography
\def\thebibliography{\DeclareRobustCommand{\VAN}[3]{##3}\VANthebibliography}

\usepackage{graphicx}	% Including figure files
\usepackage{amsmath}	% Advanced maths commands
\usepackage{bm}
\title[Accretion rates of COs in AGN discs]{Accretion rates of stellar-mass compact objects embedded in AGN discs}

\author[C.-L. Jiao et al.]{
Cheng-Liang Jiao,$^{1,2,3}$\thanks{E-mail: jiaocl@ynao.ac.cn (CLJ)}
Liying Zhu,$^{1,4,3}$
Er-gang Zhao$^{1,3}$
and Jia Zhang$^{1,3}$
\\
$^{1}$Yunnan Observatories, Chinese Academy of Sciences, 396 Yangfangwang, Guandu District, Kunming, 650216, P. R. China\\
$^{2}$Center for Astronomical Mega-Science, Chinese Academy of Sciences, 20A Datun Road, Chaoyang District, Beijing, 100012, P. R. China\\
$^{3}$Key Laboratory for the Structure and Evolution of Celestial Objects, Chinese Academy of Sciences, 396 Yangfangwang, Guandu District, Kunming, 650216, P. R. China\\
$^{4}$University of Chinese Academy of Sciences, No. 1 Yanqihu East Road, Huairou District, 101408, Beijing, P. R. China
}

\date{Accepted XXX. Received YYY; in original form ZZZ}

\pubyear{\the\year{}}

\begin{document}
\label{firstpage}
\pagerange{\pageref{firstpage}--\pageref{lastpage}}
\maketitle

\begin{abstract}
Stellar-mass compact objects (COs) embedded in active galactic nucleus (AGN) discs are commonly assumed to accrete via Bondi or Bondi-Hoyle-Lyttleton (BHL) prescriptions, neglecting gas angular momentum.
We show that differential rotation in AGN discs can impart non-negligible angular momentum, in which case accretion proceeds through a viscous disc rather than Bondi/BHL flow.
Our model provides a new framework estimating the CO accretion rate as $\dot{M}_\mathrm{CO} = \min\{\dot{M}_\mathrm{vis}, \dot{M}_\mathrm{BHL}\}$, where the viscous rate $\dot{M}_\mathrm{vis}$ accounts for gas--CO relative motion decomposed into a local gradient term (due to differential rotation) and bulk motion (from differing orbital parameters). 
This rate can be expressed as $\dot{M}_\mathrm{vis} = \alpha \xi (r_\mathrm{H}/r_\mathrm{BHL})^3\dot{M}_\mathrm{BHL}$, where $\xi$ is a coefficient of order unity. It can also be approximately scaled to the global AGN accretion rate as $\dot{M}_\mathrm{vis} \propto \dot{M}_1$, with the scaling coefficients in both forms determined by the specific dynamical configuration.
The accretion is viscosity-limited when $q > [\alpha \xi(1+\mathcal{M}^2)^{3}/3]^{1/2} h^3$, where $q$ is the mass ratio between the CO and the supermassive black hole, $\alpha$ the viscosity parameter, $\mathcal{M}$ the Mach number of the bulk relative motion, and $h$ the aspect ratio of the AGN disc. 
In thin AGN discs this condition is satisfied for most stellar-mass or more massive COs.
Our framework also naturally allows for the inclusion of established outflow corrections, thereby enabling a more realistic treatment of super-Eddington flows. 
Our formulation thus improves upon Bondi/BHL prescriptions and offers a more physically motivated basis for studying CO evolution in AGN environments.
\end{abstract}

\begin{keywords}
accretion, accretion discs -- galaxies: active -- quasars: supermassive black holes -- stars: black holes
\end{keywords}

%%%%%%%%%%%%%%%%%%%%%%%%%%%%%%%%%%%%%%%%%%%%%%%%%%

%%%%%%%%%%%%%%%%% BODY OF PAPER %%%%%%%%%%%%%%%%%%

\section{Introduction}\label{intro}

Active Galactic Nuclei (AGNs) are among the most luminous and energetic regions in the universe, powered by supermassive black holes (SMBHs) at the centre of galaxies. These SMBHs are surrounded by accretion discs--flattened structures of gas and dust spiraling inward due to gravitational forces. Within these discs, stellar-mass compact objects (COs), such as black holes (BHs), neutron stars, or white dwarfs with masses ranging from a few to tens of solar masses, can exist. These objects may form through gravitational instabilities in the disc or be captured from the surrounding nuclear star cluster \citep[][and references therein]{McK2012,Bartos2017,Tagawa2023}.
AGN discs provide a unique environment where dense gas, strong gravitational fields, and dynamical interactions converge to shape the evolution of COs in ways distinct from isolated systems or other dynamical channels. First, AGN discs act as ‘cosmic factories’ for producing massive binary black hole (BBH) mergers, including events in the pair-instability mass gap (e.g., GW190521; \citealt{Abbott2020}). The gas-rich environment facilitates rapid mass growth of COs via hyper-Eddington accretion \citep{McK2012,Li2021}, enabling hierarchical mergers that may form intermediate-mass BHs \citep{Yang2019,Frag2022}. Such mergers are challenging to explain through isolated binary evolution or dynamical channels in globular clusters, making AGN discs a compelling alternative pathway \citep{Bartos2017,Sam2022}.
Second, the interaction between COs and the AGN disc gas drives unique dynamical processes. Gas dynamical friction and tidal torques efficiently dissipate orbital energy, allowing COs to migrate radially and form binaries through close encounters \citep{LiR2023,DeL2023}. These mechanisms are critical for overcoming the ‘final au problem’ in BBH formation, where traditional channels struggle to shrink orbits to separations where gravitational wave (GW) emission dominates \citep{Stone2017}. Numerical simulations reveal that retrograde binaries or systems with non-isothermal gas equations of state experience enhanced orbital shrinkage due to asymmetric torques from circum-binary discs \citep{Li2022b,Dem2022}. Furthermore, AGN discs can imprint distinct eccentricities and spin-orbit misalignments on merging BBHs, serving as observational fingerprints to distinguish the AGN disc channel from others \citep{Sam2022,Li2022b}.
Third, accretion onto COs in AGN discs generates unique electromagnetic (EM) signals. Hyper-Eddington accretion produces luminous X-ray flares, quasi-periodic eruptions (QPEs), and transient outflows, which are potential counterparts to GW events \citep{Zhu2021,Linial2023}. For instance, accretion-induced shocks or tidal disruption events during CO--disc interactions may explain observed QPEs \citep{ZC2024}. Additionally, the interaction of merger remnants with the AGN disc could produce late-time EM afterglows, aiding in source localization \citep{Kimura2021,Perna2021}. These EM signatures, combined with GW detections, enable multi-messenger tests of the contribution of AGN disc channel to the GW population \citep{Bartos2017,Graham2020}.
Finally, the dynamics and accretion of COs embedded in AGN discs significantly shape large-scale astrophysical processes. CO--disc interactions alter disc structure, potentially resolving discrepancies in observed AGN disc sizes or metal abundances \citep{Cant2021,WM2023,ZS2024}. Stellar-mass COs may also trigger episodic AGN activity through feedback or disc fragmentation \citep{Gil2022,Tagawa2023b}. Moreover, AGN discs serve as laboratories for studying extreme accretion physics, such as super-Eddington flows and radiation feedback, which are difficult to probe in other environments \citep{Li2021,CRD2023}.

In summary, studying COs embedded in AGN discs bridges GW astronomy, high-energy astrophysics, and galaxy evolution. It addresses key questions about the origins of massive BBH mergers, provides testable predictions for multi-messenger observations, and deepens our understanding of accretion physics in extreme environments. These advances depend critically on understanding the dynamical evolution and accretion processes of COs in AGN discs. The accretion rate of the CO is a key parameter in such studies, significantly influencing the mass and spin evolution of the CO, the EM and GW signals associated with its activity, and its dynamical interaction with the AGN disc.
In previous works, the Bondi \citep{Bondi1952,Frank2002} or Bondi-Hoyle-Lyttleton (BHL; \citealt{HL1939,Shima1985,Edgar2004}) accretion rate is widely adopted to estimate the accretion rate of the CO at the outer boundary \citep[e.g.,][]{Kocsis2011,Wang2021a,Wang2021b,Pan2021, CRD2023, Liu2024,Zhang2024}. However, standard AGN disc models, such as the Shakura-Sunyaev disc (SSD; \citealt{SS73}), the marginally self-gravitating quasi-stellar object disc \citep{Sirko2003}, and the radiation pressure-supported starburst disc \citep{Thompson2005}, all assume Keplerian rotation around the SMBH. Assuming that the CO also follows a circular Keplerian orbit, differential rotation induces relative motion between the disc gas and the CO.
The relative velocity scales linearly with the gas--CO distance $r$.
Since the Keplerian velocity around the CO scales as $r^{-1/2}$, the relative velocity will eventually reach the local Keplerian value at a specific radius. Newly accreted gas past this radius has Keplerian rotation with respect to the CO and is rotationally supported rather than pressure supported. 
Consequently, the classical Bondi and BHL accretion models no longer apply, as they neglect angular momentum effects \citep{Bondi1952, Shima1985}.
The accretion flow in this scenario should be a viscous accretion disc, where gas is accreted gradually as viscous dissipation transports angular momentum outward (see section \ref{accretion} for details), rather than a Bondi or BHL accretion flow where gravity is resisted solely by pressure gradients and the gas almost free-falls in the inner regions.

\cite{Kocsis2011} considered the relative motion between the CO and the AGN disc gas. However, they treated the relative velocity as a constant value calculated at the Hill radius ($r_\mathrm{H}$) of the CO and estimated the accretion rate using the BHL rate, neglecting rotational support. Various subsequent studies (e.g., \citealt{Pan2021,CRD2023,Zhang2024}) followed this treatment and further treated the specific angular momentum calculated at $r_\mathrm{rel}=\min\{r_\mathrm{H}, r_\mathrm{BHL,prev}\}$ (where $r_\mathrm{BHL,prev}$ denotes the BHL radius as adopted in these previous studies; this differs from the classical BHL radius as discussed below) as a constant. 
In their models, the accretion flow circularises at an inner radius $r_\mathrm{cir}$ where this constant specific angular momentum matches the local Keplerian value. Within $r_\mathrm{cir}$, the accretion flow was modelled as a viscous accretion disc, while beyond this radius, the effects of angular momentum were ignored and the BHL accretion rate was imposed as a boundary condition at $r_\mathrm{cir}$. However, the relative velocity is not constant and actually scales linearly with $r$, such that the specific angular momentum with respect to the CO scales as $r^2$.
Consequently, their assertion that disc gas at larger distances from the CO undergoes sub-Keplerian rotation, which circularises at smaller radii, is incorrect. 
In reality, gas at larger distances exhibits super-Keplerian rotation, and a natural Keplerian radius exists from which gas is accreted via viscous dissipation of angular momentum.
This radius typically lies close to the Hill radius (see sections~\ref{accretion} and \ref{newsec} for details).
The estimation of the accretion rate at the Hill radius, which is commonly adopted as the effective outer boundary of the CO’s accretion region \citep[e.g.,][]{Good2004,Stone2017}, therefore requires explicit consideration of angular momentum effects.
Moreover, the relative motion induced by differential rotation is highly position-dependent and reverses its direction between the sides of the CO facing toward and away from the SMBH. Therefore, it should not be treated as the bulk inflow assumed in the classical BHL model.
In particular, when the CO and the AGN disc share the same orbital parameters, the relative velocity vanishes at the CO’s location, rendering the use of the BHL prescription with a non-zero relative velocity unsuitable.

In this paper, we explicitly account for the angular momentum of the disc gas and derive a new formula for estimating the CO accretion rate at the Hill radius within a viscous accretion disc framework. 
The relative motion between the CO and the surrounding gas is treated consistently by separating it into two components: a gradient term due to differential rotation and bulk motion arising from differences in orbital parameters. The gradient term carries net angular momentum relative to the CO, leading to the formation of a viscous accretion disc. Our formulation therefore improves upon the Bondi and BHL prescriptions for CO accretion in AGN discs, providing a more physically consistent description.

It is worth noting that several works have adopted the Eddington accretion rate \citep[e.g.,][]{McK2020b,Tagawa2021,WM2023}, defined as $\dot{M}_\mathrm{Edd}=L_\mathrm{Edd2}/(\eta c^2)$ (where $L_\mathrm{Edd2}$ is the Eddington luminosity of the CO and $\eta$ is the radiative efficiency), or the smaller value between this and the BHL rate \citep[e.g.,][]{Tagawa2020a,Tagawa2020b,Gil2022,EM2025}. 
The Eddington accretion rate adopted in these works is on the order of $10L_\mathrm{Edd2}/c^2$, depending on the choice of $\eta$. We note that the Eddington rate can be interpreted as the final accretion rate onto the CO when radiation force is considered. Theoretically speaking, disc accretion, unlike spherical accretion, permits super-Eddington luminosities \citep{Kato2008,Yuya2016} due to anisotropic emission. Furthermore, photon trapping \citep{Katz1977,Begelman1978} can significantly reduce radiative efficiency, so that hyper-Eddington accretion rates ($\gtrsim 5000L_\mathrm{Edd2}/c^2$) are plausible even when accretion is limited by the emergent luminosity \citep{Kohei2016,Yuya2016}. 
Consequently, the application of the standard Eddington limit to accreting COs embedded in AGN discs requires careful reassessment. 
Due to anisotropic emission in an accretion disc, the radiation feedback of high luminosity is more likely in the form of outflows, driven by the strong radiation force.
Our results provide essential outer boundary conditions (OBCs) for further studies incorporating these feedback effects.

The relative motion between the CO and surrounding gas in the absence of CO gravity also requires a detailed analysis, as it provides either initial conditions for time-dependent simulations or boundary conditions for steady models when CO gravity is included.
We have mentioned that many previous works calculated the relative velocity as a constant. Even in this simplified treatment, the formula for the relative velocity has been controversial. For example, \cite{Kocsis2011} and \cite{Pan2021} used $v_\mathrm{rel}=3\Omega_2 r_\mathrm{rel}/2$ (where $\Omega_2$ denotes the angular velocity of the CO relative to the SMBH), whereas \cite{CRD2023} and \cite{Zhang2024} used $v_\mathrm{rel}=\Omega_2 r_\mathrm{rel}/2$. 
We show in section \ref{rel_motion} that these formulae actually apply in different reference frames: the former applies in the non-inertial frame corotating with the CO, while the latter applies in the rest frame of the SMBH. These previous works also lack proper force analyses in their respective frames, which we present in this paper.

The paper is arranged as follows. Section \ref{rel_motion} analyses the relative motion between the CO and AGN disc gas in the absence of CO gravity, establishing the OBCs for CO accretion when its gravity is included. Building upon these boundary conditions, section \ref{accretion} derives an approximate formula for the CO accretion rate in the circular Keplerian case. 
Section~\ref{newsec} extends the formulation to the more general case.
The results are summarised and discussed in section \ref{summary}.

%%%%%%%%%%%%%%%%%%%%%%%%%%%%%%%%%%%%%%%%%%%%%%%%%%%%%%%
\section{Relative Motion Between Stellar-mass Compact Objects and AGN Disc Gas: Neglecting Compact Object Gravity}
    \label{rel_motion}

This section analyses the relative motion between the CO and the surrounding gas, neglecting CO gravity. This configuration provides either initial conditions for time-dependent simulations or boundary conditions for steady models when CO gravity is subsequently included.
We assume that both the AGN disc and the CO follow circular, corotating Keplerian orbits around the SMBH, with the orbital plane of the CO aligned with the equatorial plane of the AGN disc (see section~\ref{newsec} for discussion of the more general case with different orbital parameters).

The relative motion can be analysed in either the rest frame or the non-inertial frame corotating with the CO.
The corotating frame offers distinct computational advantages for studying accretion onto the CO. In this frame, the gravitational influence of the SMBH manifests as a tidal force and we need only consider the motion of accreted gas relative to the CO, rather than the combination of its large-scale circular motion around the SMBH and its relative motion.
A standard characteristic scale in such analyses is the Hill radius,
\begin{equation}\label{rH}
r_\mathrm{H} \equiv \left( \frac{M_2}{3M_1} \right)^{1/3} R_2,
\end{equation}
where $M_1$ and $M_2$ are the masses of the SMBH and the CO respectively, and $R_2$ is their separation. We note that $r_\mathrm{H}$ is inherently defined in the corotating frame, as it marks the boundary where the tidal force of the SMBH, rather than its gravitational force,  equals the gravitational force of the CO. In fact, most previous works explicitly or implicitly adopted the corotating frame, which can be inferred from whether they included the gravitational force of the SMBH in their equations. However, since several recent works (e.g., \citealt{CRD2023,Zhang2024}) calculated relative velocities in the rest frame, we provide a detailed analysis of the relative motion in this frame for completeness in section \ref{rel_rest}, followed by the analysis in the corotating frame in section \ref{rel_coro}.

\subsection{Relative motion in the rest frame}\label{rel_rest}
In the rest frame of the SMBH, we define a cylindrical coordinate system ($R$, $\Phi$, $Z$) centred on the SMBH, whose equatorial plane coincides with that of the AGN disc. The coordinates of the CO are thus ($R_2$, $\Phi_2$, 0).
Assuming that the velocities of the CO and the disc gas can be described by the same function $\bm{v}(R, \Phi, Z)$, the relative velocity in the vicinity of the CO is given by
\begin{equation}\label{}
\Delta \bm{v}=\Delta x^i \nabla_i\bm{v},
\end{equation}
where $\Delta x^i$ ($i=1$, 2, 3) represents small variations of the coordinates and $\nabla_i \bm{v}$ represents the covariant derivative of $\bm{v}$.  Evaluating this expression yields
\begin{equation}\label{eq_dv1}
\Delta v_R=\frac{\partial v_R}{\partial R} \Delta R + \frac{\partial v_R}{\partial \Phi} \Delta \Phi + \frac{\partial v_R}{\partial Z} \Delta Z
- v_\Phi \Delta \Phi,
\end{equation}
\begin{equation}\label{eq_dv2}
\Delta v_\Phi=\frac{\partial v_\Phi}{\partial R} \Delta R + \frac{\partial v_\Phi}{\partial \Phi} \Delta \Phi + \frac{\partial v_\Phi}{\partial Z} \Delta Z
+ v_R \Delta \Phi,
\end{equation}
\begin{equation}\label{eq_dv3}
\Delta v_Z= \frac{\partial v_Z}{\partial R} \Delta R + \frac{\partial v_Z}{\partial \Phi} \Delta \Phi + \frac{\partial v_Z}{\partial Z} \Delta Z.
\end{equation}
The last terms in equations (\ref{eq_dv1}) and (\ref{eq_dv2}) arise from the curvature of the cylindrical coordinate system.

\begin{figure}
\includegraphics[width=\columnwidth]{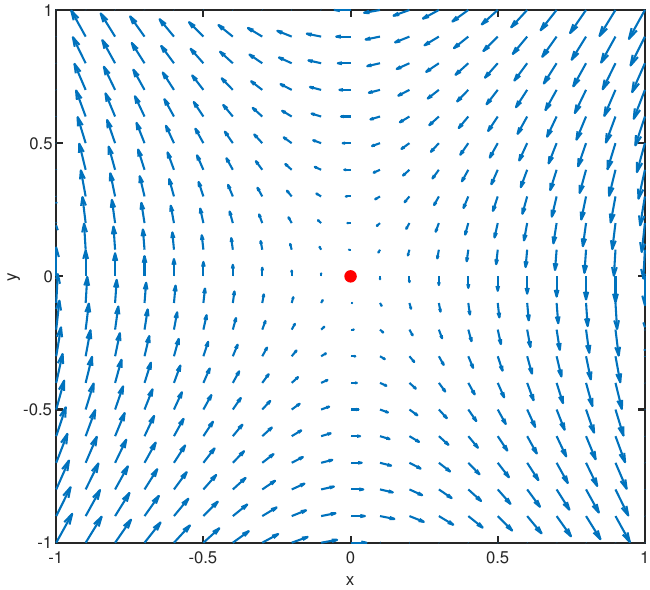}
\caption{Velocity field of the surrounding gas relative to the CO in the rest frame. The central red dot represents the CO. The unit of length is much smaller than $R_2$. The $x$- and $y$-directions correspond to $\bm{e}_\Phi$ and $\bm{e}_R$ at the CO. }
\label{fig1}
\end{figure}

Since both the CO and the disc gas follow Keplerian orbits, their azimuthal velocity is given by the same expression, $v_\Phi=\sqrt{GM_1/R}$. In standard AGN disc models \citep{SS73,Sirko2003,Thompson2005}, both $v_R$ and $v_Z$ are typically negligible compared to $v_\Phi$. The CO orbits on the equatorial plane of the AGN disc, so it has no vertical motion. Furthermore, its radial velocity due to disc-induced migration is also negligible compared to $v_\Phi$. Consequently, both the disc gas and the CO satisfy $v_Z \approx 0$ and $v_R \approx 0$. 
Substituting these relations into equations (\ref{eq_dv1})--(\ref{eq_dv3}), we derive
\begin{align}
\Delta v_R &\approx - \Omega_2 R \Delta \Phi, \label{rest_dv1}\\
\Delta v_\Phi &\approx -\frac{\Omega_2}{2} \Delta R, \label{rest_dv2}\\
\Delta v_Z &\approx 0, \label{rest_dv3}
\end{align}
where $\Omega_2=\sqrt{GM_1/R_2^3}$ is the angular velocity of the CO with respect to the SMBH. Equation (\ref{rest_dv2}) corresponds to the relative velocity formula adopted by \citet{CRD2023} and \citet{Zhang2024}. However, the rotational motion not only generates the differential velocity in the azimuthal direction (equation (\ref{rest_dv2})), but also induces radial relative motion (equation (\ref{rest_dv1})).
This is clearly illustrated in Fig. \ref{fig1}, which displays the velocity field of the surrounding gas relative to the CO in the rest frame. When viewed face-on (clockwise orbital motion assumed), two distinct components emerge: the $x$-component is produced by differential rotation, where gas at larger radii moves slower than the CO, resulting in net motion opposite to the orbital direction; the $y$-component is generated by orbital curvature, where the azimuthal unit vector $\bm{e}_\Phi$ along the CO orbit diverges geometrically ($\bm{e}_\Phi$ to the left of the CO points slightly upward, while that to the right points slightly downward).

The forces acting on the disc gas in the rest frame comprise the gravitational forces of both the SMBH and the CO. Critically, the gravitational influence of the SMBH must be retained even within $r_\mathrm{H}$, as the motion of the surrounding gas is the combination of its large-scale orbital motion around the SMBH and its local motion relative to the CO.

\subsection{Relative motion in the corotating frame}\label{rel_coro}

\begin{figure}
\includegraphics[width=\columnwidth]{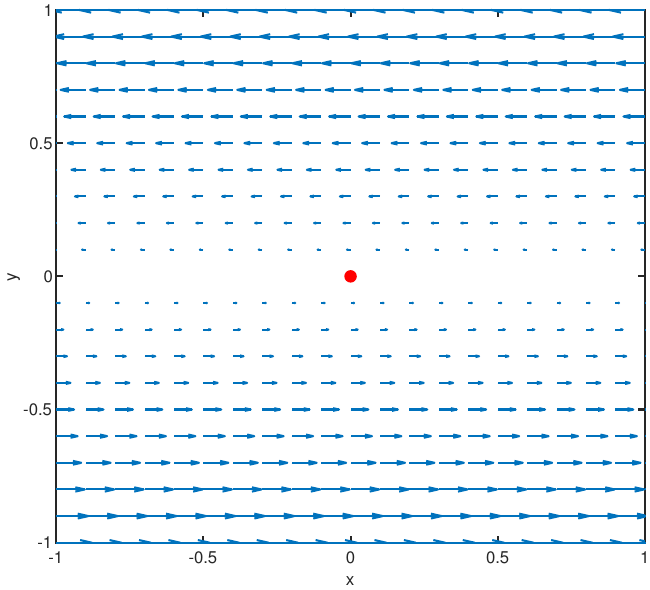}
\caption{Velocity field of the surrounding gas relative to the CO in the corotating frame. The central red dot represents the CO. The unit of length is much smaller than $R_2$. The $x$- and $y$-directions correspond to $\bm{e}_\Phi$ and $\bm{e}_R$ at the CO. }
\label{fig2}
\end{figure}

We now consider the non-inertial frame corotating with the CO at angular velocity $\Omega_2$. In this frame, the velocity components are $v_\Phi=\sqrt{GM_1/R}-\Omega_2R$, $v_Z \approx 0$, and $v_R \approx 0$, where $\Omega_2$ is treated as a constant. In the vicinity of the CO ($R \approx R_2$), substituting these relations into equations (\ref{eq_dv1})--(\ref{eq_dv3}) yields
\begin{align}
\Delta v_R &\approx 0, \label{coro_dv1}\\
\Delta v_\Phi &\approx -\frac{3\Omega_2}{2} \Delta R, \label{coro_dv2}\\
\Delta v_Z &\approx 0. \label{coro_dv3}
\end{align}
The velocity field of the surrounding gas relative to the CO in the corotating frame is displayed in Fig. \ref{fig2}. The forces acting on the disc gas in the corotating frame include the non-inertial forces (centrifugal and Coriolis), in addition to the gravitational forces of the SMBH and the CO. These forces will be investigated in detail in section \ref{accretion}.

%%%%%%%%%%%%%%%%%%%%%%%%%%%%%%%%%%%%%%%%%%%%%%%%%%%%%%%
\section{Accretion onto stellar-mass compact objects: the circular Keplerian case}\label{accretion}

We investigate the accretion onto the CO in the corotating frame, assuming that both the AGN disc and the CO follow coplanar circular Keplerian orbits around the SMBH as presented in section~\ref{rel_motion}.
Two coordinate systems are employed: Cartesian ($x$, $y$, $z$) and cylindrical ($r$, $\phi$, $z$), both centred on the CO. 
The $x$- and $y$-directions in the Cartesian system correspond to $\bm{e}_\Phi$ and $\bm{e}_R$ of the global disc coordinates at the CO, respectively, such that $\bm{e}_z$ is opposite to $\bm{e}_Z$ to maintain right-handedness. The cylindrical system shares the same $z$-axis as the Cartesian one, while $\phi$ denotes the azimuthal angle measured from $\bm{e}_y$. The coordinate transformations are given by
\begin{align}
x &= -r\sin\phi, \\
y &= r\cos\phi.
\end{align}

As the corotating frame is a non-inertial frame, in addition to the gravitational forces of the CO and the SMBH, we also need to consider non-inertial forces, including Coriolis and centrifugal forces (the Euler force is zero since $\bm{\Omega}_2$ is constant in the time interval of interest). The sum of the centrifugal force and the gravitational force of the SMBH is also known as the tidal force \citep{Ormel2015a}, which is
\begin{equation}\label{f_tid}
\bm{F}_\mathrm{tid}= \left[ \Omega_2^2(R_2+y)-\frac{GM_1}{(R_2+y)^2} \right] \bm{e}_y \approx 3\Omega_2^2y \bm{e}_y,
\end{equation}
where the approximation holds for $|y| \ll R_2$.
The Coriolis force is
\begin{equation}\label{f_cor}
    \bm{F}_\mathrm{cor}=-2\bm{\Omega}_2 \times \bm{v}
    =2\Omega_2 \bm{e}_z\times \bm{v},
\end{equation}
where $\bm{\Omega}_2=-\Omega_2 \bm{e}_z$ since $\bm{e}_z=-\bm{e}_Z$.
The initial relative velocity components are specified by equations (\ref{coro_dv1})--(\ref{coro_dv3}). Expressed in the local Cartesian coordinate system, these become
\begin{align}
v_x &\approx -\frac{3\Omega_2}{2} y, \label{coro_v1} \\
v_y &\approx 0, \label{coro_v2} \\
v_z &\approx 0. \label{coro_v3}
\end{align}
The Coriolis force is thus $\bm{F}_\mathrm{cor}=-3\Omega_2^2y \bm{e}_y$, which exactly balances the tidal force in the initial configuration. 
Upon considering the gravitational force of the CO, the gas motion diverges from the initial configuration. Consequently, the Coriolis force ceases to fully balance the tidal force, though it retains partial counteracting capability.

According to equation (\ref{coro_v1}), the rotation of the surrounding gas relative to the CO is opposite to the orbital motion of the CO. The angular velocity in the initial configuration is given by
\begin{equation}\label{coro_omega}
\bm{\omega}_\mathrm{ini}= \frac{3}{2}\Omega_2 \cos^2{\phi} \bm{e}_z,
\end{equation}
which shows no radial dependence.
As the Keplerian angular velocity around the CO is $\omega_\mathrm{K} = \sqrt{GM_2/r^3}$, the disc gas reaches Keplerian rotation at a characteristic radius. For $\phi=0$, this Keplerian radius is
\begin{equation}\label{eq_rK}
r_\mathrm{K} = \left(\frac{4M_2}{9M_1}\right)^{\frac{1}{3}} R_2=\left(\frac{4}{3}\right)^{\frac{1}{3}} r_\mathrm{H}.
\end{equation}
For $\phi \neq 0$, the radius scales as $(\cos\phi)^{-4/3} r_\mathrm{K}$.

For simplicity, in the following derivation we focus on the gas along the SMBH--CO line (i.e., at $\phi = 0$ or $\pi$), where the relative motion reaches its maximum value at a given radius $r$. 
Following common practice in the literature \citep[e.g.,][]{Good2004,Stone2017}, we adopt the Hill radius $r_\mathrm{H}$ as the effective outer boundary of the CO's accretion region.
Beyond $r_\mathrm{H}$ the tidal force of the central SMBH dominates, and we use the unperturbed gas properties of the AGN disc presented in section~\ref{rel_coro} as the boundary conditions there.
Within $r_\mathrm{H}$ the gravitational force of the CO (denoted $g_2$) prevails, and we neglect tidal and Coriolis forces.\footnote{
By definition, the tidal force becomes less than $g_2$ within $r_\mathrm{H}$. The Coriolis force depends on the velocity of the gas; assuming a Keplerian rotation around the CO, it becomes less than $g_2$ within $(3/4)^{1/3}r_\mathrm{H}\approx 0.91r_\mathrm{H}$.
Given that the tidal force partially offsets the Coriolis force, we may neglect both forces in an order-of-magnitude estimate for $r \leq r_\mathrm{H}$.}
Since $\bm{\omega}_\mathrm{ini}$ exhibits no radial dependence, the initial configuration lacks differential rotation and therefore experiences no viscous dissipation in the corotating frame.
Within $r_\mathrm{H}$, $\bm{\omega}_\mathrm{ini}$ is sub-Keplerian ($r_\mathrm{H} < r_\mathrm{K}$), causing the gas to move inward under gravity.
With no viscous dissipation, the gas would eventually encounter a centrifugal barrier. However, accretion is sustained by two mechanisms: (1) depletion of gas closest to the accretor creates a pressure gradient that drives adjacent gas inward across the centrifugal barrier; (2) as gas flows inward, its angular velocity increases, establishing differential rotation relative to outer gas and consequently enabling viscous processes (or equivalent processes such as magneto-rotational instability). It is thus plausible that eventually a steady viscous accretion disc forms around the CO, which rotates with angular velocities higher than $3\Omega_2/2$ inside $r_\mathrm{H}$, facilitating viscous dissipation.

To derive an analytically tractable approximation for the accretion rate, we assume that the surrounding gas eventually forms a steady, axisymmetric accretion disc around the CO, with outer radius $r_\mathrm{out}=r_\mathrm{H}$.
The OBCs for the density, isothermal sound speed, and angular velocity of the gas at $r_\mathrm{out}$ are the unperturbed values, i.e., the local density and sound speed of the AGN disc gas (denoted $\rho_\mathrm{d}$ and $c_\mathrm{sd}$, respectively) and the initial angular velocity $3\Omega_2/2$.
While the gas rotation is slightly sub-Keplerian at $r_\mathrm{out}=r_\mathrm{H}$, recent numerical simulations \citep[e.g.,][]{Bu2014} have shown that, for a steady accretion flow,
the angular momentum profile remains smooth and does not develop a distinct circularisation radius, even when the angular momentum at the outer boundary is significantly sub-Keplerian. 
We therefore proceed to calculate the accretion rate of the disc around the CO without invoking an additional circularisation process.
The angular momentum equation is written as
\begin{equation}\label{eq_motion_phi}
\frac{\partial }{\partial r}(-\dot{M} \omega r^2) = \frac{\partial \mathcal{G}}{\partial r},
\end{equation}
where $\mathcal{G}$ represents the viscous torque
\begin{equation}\label{eq_torque}
\mathcal{G} = 4\pi r^2 H t_{r\phi}.
\end{equation}
Here $t_{r\phi}$ denotes the $r\phi$-component of the viscous stress tensor.
The integration constant of equation (\ref{eq_motion_phi}) is on the order of $\dot{M}l_\mathrm{in}$, where $l_\mathrm{in}$ is the specific angular momentum at the inner boundary of the disc $r_\mathrm{in}$.
For neutron stars and white dwarfs, $r_\mathrm{in}$ resides within the boundary layer adjacent to the surface of the CO \citep{Frank2002}. For BHs, $r_\mathrm{in}$ can be estimated by the innermost stable circular orbit (ISCO). Thus, $r_\mathrm{in} \ll r_\mathrm{out}$ for the CO. Given that specific angular momentum scales as $r^{1/2}$, $l_\mathrm{in}$ is negligible compared to its value at $r_\mathrm{out}$.
Consequently, we obtain
\begin{equation}\label{eq_motion_phi_approx}
     \mathcal{G} = -\dot{M} \omega r^2
\end{equation}
at $r=r_\mathrm{out}$.
We adopt the $\alpha$-prescription of viscosity \citep{SS73}, $t_{r\phi}=-\alpha p=-\alpha \rho c_\mathrm{s}^2$, and the relation $H = c_\mathrm{s}/\omega_\mathrm{K}$ for a viscous accretion disc. 
From equations~\eqref{eq_torque} and \eqref{eq_motion_phi_approx}, we derive
\begin{equation}\label{eq_mdotH}
    \dot{M}_\mathrm{vis} = \frac{4\alpha \pi \rho_\mathrm{d}c_\mathrm{sd}^3}{\omega(r_\mathrm{out}) \cdot \omega_\mathrm{K}(r_\mathrm{out})}.
\end{equation}
Here, $\omega(r_\mathrm{out})=3\Omega_2/2$ is a constant, and $\omega_\mathrm{K}(r_\mathrm{out})=\sqrt{GM_2/r_\mathrm{H}^3}=\sqrt{3}\Omega_2$, such that
\begin{equation}\label{eq_mdotH2}
    \dot{M}_\mathrm{vis,H}
    = \frac{4\alpha \pi \rho_\mathrm{d} c_\mathrm{sd}^3}{3\sqrt{3} GM_1/(2R_2^3)}
    = \frac{8\alpha \pi \rho_\mathrm{d}c_\mathrm{sd}^3 r_\mathrm{H}^3}{\sqrt{3}GM_2},
\end{equation}
where we have used equation~\eqref{rH} to eliminate $R_2$ and $M_1$ in favour of $r_\mathrm{H}$ and $M_2$.
For comparison, the Bondi accretion rate can be written as\footnote{
Here we define the Bondi rate in accordance with the BHL rate \citep{Edgar2004}, neglecting the coefficient dependent on the adiabatic index $\gamma$ \citep{Bondi1952,Frank2002}.
}
\begin{equation}\label{Bondi1}
\dot{M}_{\mathrm{B}}
= \frac{4\pi \rho_\mathrm{d} c_\mathrm{sd}^3 
r_\mathrm{B}^3}{G M_2},
\end{equation}
where $r_\mathrm{B}$ is the Bondi radius of the CO, defined as
\begin{equation}\label{eq_rB}
r_\mathrm{B} \equiv \frac{GM_2}{c_\mathrm{sd}^2}.
\end{equation}
We can therefore rewrite equation~\eqref{eq_mdotH2} as
\begin{equation}\label{eq_mdotvis_rH}
    \dot{M}_\mathrm{vis}
    = \alpha \frac{2}{\sqrt{3}} \left(\frac{r_\mathrm{H}}{r_\mathrm{B}}\right)^3\dot{M}_\mathrm{B}.
\end{equation}
We note that rotational support of the gas should suppress accretion, implying $\dot{M}_\mathrm{vis} < \dot{M}_\mathrm{B}$. For $r_\mathrm{H} > r_\mathrm{B}$, equation (\ref{eq_mdotvis_rH}) may yield $\dot{M}_\mathrm{vis} \geq \dot{M}_\mathrm{B}$. However, the magnitude of relative velocity at $r_\mathrm{B}$ would be much smaller than the sound speed in this case, such that relative motion at $r_\mathrm{B}$ can be neglected and we can simply use the Bondi rate.

\begin{figure*}
    \centering
    \includegraphics[width=\textwidth]{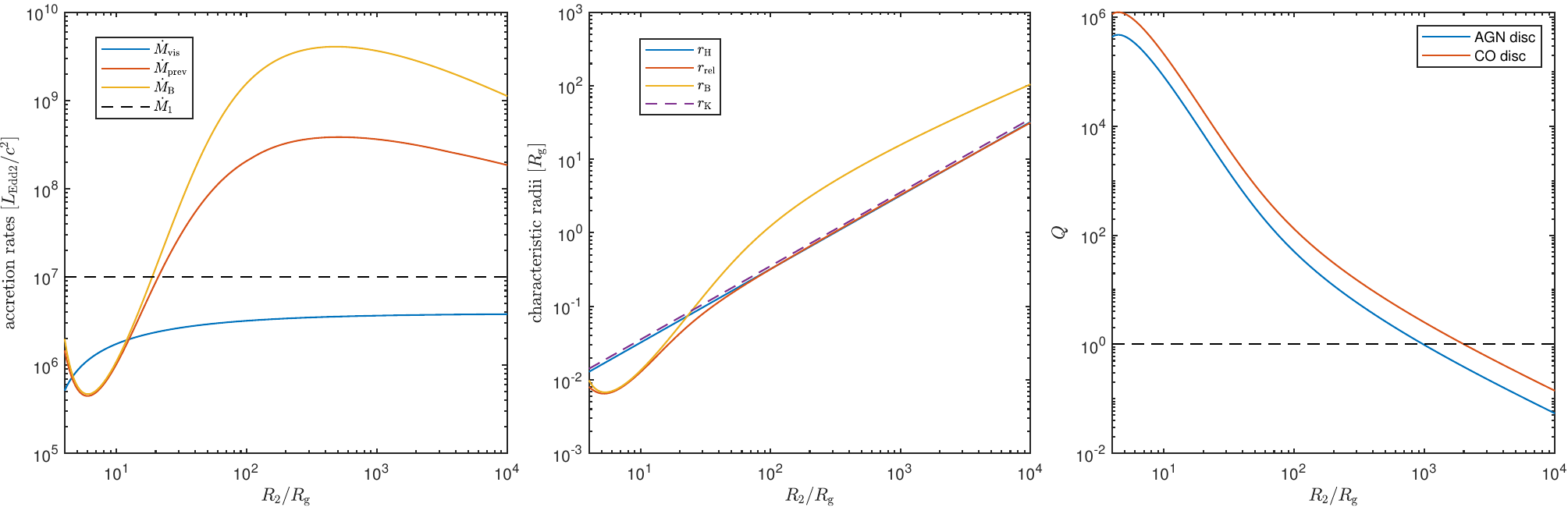}
    \caption{
        Profiles of 
        accretion rates ($\dot{M}_\mathrm{vis}$, $\dot{M}_\mathrm{prev}$, $\dot{M}_\mathrm{B}$), 
        characteristic radii ($r_\mathrm{H}$, $r_\mathrm{rel}$, $r_\mathrm{B}$, $r_\mathrm{K}$),
        and Toomre $Q$ as functions of the SMBH--CO separation $R_2$.
        The accretion rates are normalised to $L_\mathrm{Edd2}/c^2$. 
		The dashed line in the left panel represents the accretion rate of the AGN disc, $\dot{M}_1$.
		The input parameters are $M_1=10^8M_\odot$, $M_2=10M_\odot$, $\alpha=0.1$, $\dot{M}_1=L_\mathrm{Edd1}/c^2$ (corresponding to $10^7L_\mathrm{Edd2}/c^2$), $\mu=0.6$, and $\kappa_\mathrm{es}=0.35$ cm$^2$ g$^{-1}$.
    }
    \label{fig3}
\end{figure*}

To demonstrate the accretion properties of the CO, we present a representative case in Fig.~\ref{fig3}.
The input parameters are $M_1=10^8M_\odot$, $M_2=10M_\odot$, $\alpha=0.1$, and the accretion rate of the SMBH $\dot{M}_1=L_\mathrm{Edd1}/c^2$, where
\begin{equation}\label{Ledd1}
    L_\mathrm{Edd1} \equiv \frac{4\pi c G M_1 }{\kappa_\mathrm{es}}
\end{equation}
is the Eddington luminosity of the SMBH, and $\kappa_\mathrm{es}$ is the electron scattering opacity.
The properties of the AGN disc are calculated by numerically solving the equations for the SSD model presented in \cite{Kato2008}, assuming solar abundances for the disc gas (mean molecular weight $\mu=0.6$ and electron scattering opacity $\kappa_\mathrm{es}=0.35$ cm$^2$ g$^{-1}$). We assume that the AGN disc and the accretion disc around the CO have the same viscosity parameter $\alpha$.
Accretion rates are normalised to $L_\mathrm{Edd2}/c^2$, where
the Eddington luminosity of the CO is
\begin{equation}\label{Ledd2}
    L_\mathrm{Edd2} \equiv \frac{4\pi c G M_2 }{\kappa_\mathrm{es}}.
\end{equation}
The SMBH--CO separation $R_2$ is expressed in units of the gravitational radius of the SMBH, $R_\mathrm{g}=2GM_1/c^2$.

The left panel of Fig.~\ref{fig3} shows the profiles of $\dot{M}_\mathrm{vis}$ and $\dot{M}_\mathrm{B}$ as functions of $R_2$. 
For comparison, we also plot the accretion rate at the outer boundary as proposed by previous works \citep[e.g.,][]{Pan2021,CRD2023,Zhang2024}, denoted $\dot{M}_\mathrm{prev}$, which is obtained from the BHL formula \citep[e.g.,][]{Edgar2004} with the relative velocity at $r_\mathrm{rel} = \min\{r_\mathrm{H}, r_\mathrm{BHL,prev}\}$ taken as the bulk velocity.
In the present scenario, the true bulk relative motion between the CO and the disc gas is zero, and the relative motion arises only from shear and curvature (see section~\ref{rel_motion}), vanishing at the CO's location.
Therefore, $\dot{M}_\mathrm{prev}$ differs from the BHL rate calculated using the true bulk relative velocity, which is employed in the general case of our model presented in section~\ref{newsec}.

In most regions of the AGN disc shown in Fig.~\ref{fig3}, $\dot{M}_\mathrm{vis}$ is orders of magnitude lower than $\dot{M}_\mathrm{B}$ and $\dot{M}_\mathrm{prev}$, with $r_\mathrm{H}$ much smaller than $r_\mathrm{B}$ and $r_\mathrm{rel} = r_\mathrm{H}$ by definition.
An exception occurs in the radiation pressure-dominated region close to the SMBH, where high sound speeds strongly suppress both $r_\mathrm{B}$ and $\dot{M}_\mathrm{B}$ (note that $\dot{M}_\mathrm{B} \propto c_\mathrm{sd}^{-3}$ when the dependence of $r_\mathrm{B}$ on $c_\mathrm{sd}$ is accounted for in equation~\eqref{Bondi1}). 
In this region, the relative velocity of disc gas at $r_\mathrm{B}$ becomes negligible compared to the sound speed, leading to $r_\mathrm{rel} \sim r_\mathrm{B}$.
Consequently, $\dot{M}_\mathrm{B}$ or $\dot{M}_\mathrm{prev}$ should be adopted, which converge due to negligible relative motion, as verified in Fig. \ref{fig3} where $\dot{M}_\mathrm{prev} \sim \dot{M}_\mathrm{B}$ when $\dot{M}_\mathrm{vis} > \dot{M}_\mathrm{B}$.
This region diminishes or even vanishes for smaller $\alpha$, since $\dot{M}_\mathrm{vis} \propto \alpha$.
Therefore, our formula applies in most of the AGN disc, except for a small region close to the SMBH ($R \lesssim 10R_\mathrm{g}$; see section~\ref{lower_q} for a discussion of the general case).

The dashed line in the left panel of Fig.~\ref{fig3} represents $\dot{M}_1$, which is constant at $10^7L_\mathrm{Edd2}/c^2$. 
Notably, both $\dot{M}_\mathrm{B}$ and $\dot{M}_\mathrm{prev}$ exceed $\dot{M}_1$ in most of the AGN disc, which is unphysical. \cite{Kocsis2011} discussed this scenario and proposed that the CO accretion rate should be $\min\{\dot{M}_\mathrm{prev}, \dot{M}_1\}$ due to limited gas supply. However, if the CO accretes at $\dot{M}_1$, then the accretion of the SMBH is effectively halted and the system is unlikely to remain steady.
By contrast, $\dot{M}_\mathrm{vis}$ is at least several times lower than $\dot{M}_1$ and its influence on the AGN disc can be neglected in order-of-magnitude estimates.

%=========这里加入AGN和CO吸积率的关系
We can derive a direct relation between $\dot{M}_\mathrm{vis}$ and $\dot{M}_1$. 
For $R_2 \gg R_\mathrm{in} \sim 3R_\mathrm{g}$, the same derivation of the viscous accretion rate also applies to the AGN disc, yielding 
\begin{equation}\label{mdot1}
    \dot{M}_1 = \frac{4\alpha \pi \rho_\mathrm{d}c_\mathrm{sd}^3}{\Omega_2^2}.
\end{equation}
Combining this with equation~\eqref{eq_mdotH}, we obtain 
\begin{equation}\label{mdot1_mdotvis}
    \dot{M}_\mathrm{vis} = \frac{2}{3\sqrt{3}} \dot{M}_1 \approx 0.38 \dot{M}_1.
\end{equation}
This behaviour is evident in Fig.~\ref{fig3}, where $\dot{M}_\mathrm{vis}$ asymptotically approaches $2/(3\sqrt{3})\,\dot{M}_1$ at large $R_2$.

The viscous accretion rate $\dot{M}_\mathrm{vis}$ shown in Fig.~\ref{fig3} reaches magnitudes of order $10^6L_\mathrm{Edd2}/c^2$.
This is a very high accretion rate, despite being much lower than the Bondi rate (typically $10^8$--$10^9L_\mathrm{Edd2}/c^2$). 
We note that $\dot{M}_\mathrm{vis}$ represents the predicted accretion rate at the outer boundary $r_\mathrm{H}$. The actual accretion rate onto the CO could be further reduced by mechanisms such as radiation feedback and outflows (see section \ref{summary}).

An accretion disc is self-gravitationally stable if it satisfies Toomre's stability criterion $Q>1$ \citep{Too1964}, where $Q$ is defined as
\begin{equation}\label{eq_Q}
    Q \equiv \frac{c_\mathrm{s} \kappa}{\pi G \Sigma}.
\end{equation}
For simplicity, we assume that the angular velocity
profiles in both the AGN disc and the disc around the CO follow
$\Omega \propto r^{-3/2}$. The epicyclic frequency is then given by
\begin{equation}\label{epi_f}
\kappa=\sqrt{\frac{2 \Omega}{r} \frac{d}{d r}\left(r^2 \Omega\right)}=\Omega.
\end{equation}
Using $H=c_\mathrm{s}/\Omega_\mathrm{K}$, $Q$ can be expressed as
\begin{equation}\label{Q_general}
Q = \frac{c_\mathrm{s} \kappa}{\pi G \Sigma} = \frac{c_\mathrm{s} \Omega}{2\pi G \rho H} = \frac{\Omega \cdot \Omega_\mathrm{K}}{2\pi G \rho}.
\end{equation}

In the scenario presented in this section, the $Q$ parameter for the AGN disc at the location of the CO is
\begin{equation}\label{Q1}
Q_1 =  \frac{\Omega_2^2}{2\pi G \rho_\mathrm{d}}.
\end{equation}
For the viscous accretion disc around the CO, the minimum $Q$ occurs at the outer boundary $r_\mathrm{out}$, denoted $Q_2$:
\begin{equation}\label{Q2_c3}
Q_2 = \frac{\omega(r_\mathrm{out}) \cdot \omega_\mathrm{K}(r_\mathrm{out})}{2\pi G \rho_\mathrm{d}} = \frac{3\sqrt{3}}{2} Q_1,
\end{equation}
where $\omega(r_\mathrm{out})=3\Omega_2/2$ and $\omega_\mathrm{K}(r_\mathrm{out})=\sqrt{3}\Omega_2$ as discussed above.
The CO accretion disc is self-gravitationally stable provided that $Q_2>1$, indicating that its stability is guaranteed whenever the ambient AGN disc at the CO’s location is stable ($Q_1>1$).
The right panel of Fig.~\ref{fig3} demonstrates this relationship, plotting profiles of $Q_1$ and $Q_2$ as functions of $R_2$.

The SSD solution for the AGN disc becomes unstable at $R_2 \gtrsim 10^3R_\mathrm{g}$. Models with additional heating mechanisms \citep{Sirko2003,Thompson2005} in these unstable regions have been proposed for the AGN disc, where the extra heating puffs up the disc and consequently lowers the density $\rho$, stabilizing the disc. As equation (\ref{Q2_c3}) remains valid for these stable models, the accretion disc around the CO retains stability in the outer regions of the AGN disc. Consequently, our formula for $\dot{M}_\mathrm{vis}$ remains applicable provided that these stable models are adopted for the AGN disc.

%%%%%%%%%%%%%%%%%%%%%%%%%%%%%%%%%%%%%%%%%%%%%%%%%%%%%%%
\section{Accretion onto stellar-mass compact objects: the general case}\label{newsec}

In realistic astrophysical scenarios, the assumptions of purely Keplerian rotation and negligible radial velocity may break down. Both the CO and the AGN disc gas can exhibit non-Keplerian rotation or possess a non-negligible radial velocity component $v_R$. For instance, the CO could be on an eccentric orbit or have a relatively large inspiral speed. The AGN disc itself might be better described by a slim disc \citep{Ab1988} or an advection-dominated accretion flow (ADAF; \citealt{NY94,NY95a}), where rotation is sub-Keplerian and radial velocity is non-negligible.

In this more general case, the disc gas possesses a bulk motion relative to the CO, $\bm{v}_\mathrm{bulk}$, defined as its velocity at the location of the CO in the non-inertial frame comoving with the CO.\footnote{This frame corotates with the CO around the SMBH and shares its instantaneous radial velocity, ensuring that the CO remains fixed at the origin.} 
The total relative velocity is then composed of $\bm{v}_\mathrm{bulk}$ and the contribution from the local velocity gradient, $\bm{v}_\mathrm{grad}$, discussed in section~\ref{rel_motion}.
If both $\bm{v}_\mathrm{grad}$ and viscosity are neglected, the setup corresponds to the classical BHL case. 
We build on this picture by incorporating the effects of $\bm{v}_\mathrm{grad}$ and viscous angular-momentum transport.

In the classical BHL model \citep[e.g.,][and references therein]{Edgar2004}, 
gas is captured and gravitationally focused toward the accretor within a cylindrical region 
defined by the accretion radius (also known as the Hoyle--Lyttleton radius or the critical impact parameter), 
$r_\mathrm{A} = 2r_\mathrm{BHL}$. 
The BHL radius is given by
\begin{equation}\label{rBHL_bulk}
r_\mathrm{BHL} = \frac{GM_2}{c_\mathrm{sd}^2+V_\mathrm{b}^2}
= \frac{GM_2}{c_\mathrm{sd}^2(1+\mathcal{M}^2)}
= \frac{r_\mathrm{B}}{1+\mathcal{M}^2},
\end{equation}
where $V_\mathrm{b} \equiv \lvert\bm v_{\rm bulk}\rvert$ is the magnitude of the bulk relative velocity, 
and $\mathcal{M} \equiv V_\mathrm{b}/c_\mathrm{sd}$ is the Mach number of the bulk relative motion. 
The corresponding accretion rate takes the canonical form
\begin{equation}\label{BHL_rate_bulk}
\dot{M}_{\mathrm{BHL}} = \frac{4\pi G^2 M_2^2 \rho_\mathrm{d}}{(c_\mathrm{sd}^2+V_\mathrm{b}^2)^{3/2}}
= \frac{4\pi G^2 M_2^2 \rho_\mathrm{d}}{c_\mathrm{sd}^3(1+\mathcal{M}^2)^{3/2}}
= \frac{\dot{M}_\mathrm{B}}{(1+\mathcal{M}^2)^{3/2}}.
\end{equation}

The classical BHL model features only bulk motion symmetric about the accretion axis, i.e., the axis aligned with the upstream flow direction and passing through the accretor (sometimes referred to as the symmetry axis or stagnation line). Due to this symmetry, the net angular momentum of the flow relative to the accretor vanishes, resulting in the formation of an accretion column.

In our scenario, only the angular momentum associated with $\bm{v}_\mathrm{bulk}$ exhibits this symmetry and cancels out when the flow converges. 
The angular momentum from $\bm{v}_\mathrm{grad}$ remains, likely leading to the formation of a viscous accretion disc around the CO. 
Consequently, we interpret $\dot{M}_\mathrm{BHL}$ as the mass capture rate, while the actual accretion rate onto the CO is limited by viscous transport of the retained angular momentum. Our subsequent analysis therefore focuses exclusively on the angular momentum contribution from $\bm{v}_\mathrm{grad}$.

An important factor controlling the accretion process is the Mach number $\mathcal{M}$, which determines the flow morphology around the CO. 
For subsonic flows ($\mathcal{M}<1$), the gas is smoothly focused toward the CO, and the OBCs for the viscous accretion disc are taken as the local disc values, $\rho_\mathrm{d}$ and $c_\mathrm{sd}$.

By contrast, for supersonic flows ($\mathcal{M}>1$),  converging streamlines collide and form a bow shock. Across the shock, the gas loses kinetic energy, becomes compressed and heated, and is redirected toward the accretion axis \citep[e.g.,][]{Edgar2004}.
Post-shock gas properties can be approximated using the Rankine--Hugoniot jump conditions (assuming inviscid, non-magnetized gas across the shock), which depend on the normal Mach number. Quantifying the post-shock state therefore requires knowledge of the orientation and geometry of the shock surface.
Numerical simulations typically find a concave shock surface, with smaller inclination angles relative to the upstream flow for higher Mach numbers, and larger angles near the accretion axis, reaching $\pi/2$ at the axis--shock intersection \citep[e.g.,][]{Edgar2004,Mel2015}.
For very high $\mathcal{M}$, however, the accretion rate is likely limited by $\dot{M}_\mathrm{BHL}$ rather than $\dot{M}_\mathrm{vis}$.
Therefore, in the parameter regime relevant to our calculations, $\mathcal{M}$ is not excessively large, and the viscous accretion disc is expected to have a smaller outer radius than $r_\mathrm{A}$.
We can thus approximate the shock surface as perpendicular to $\bm{v}_\mathrm{bulk}$.
Consequently, the post-shock quantities can be expressed in terms of dimensionless shock jump factors, defined as
\begin{equation}
\epsilon_\rho \equiv \frac{\rho'}{\rho_\mathrm{d}}, \quad
\epsilon_v \equiv \frac{v_\mathrm{n}'}{v_\mathrm{n}}, \quad
\epsilon_p \equiv \frac{p'}{p_\mathrm{d}}, \quad
\epsilon_{c_\mathrm{s}} \equiv \frac{c_\mathrm{s}'}{c_\mathrm{sd}},
\end{equation}
where $v_\mathrm{n}$ is the velocity component normal to the shock.
Tangential velocity components are continuous across the shock.
Using the Rankine--Hugoniot jump conditions across the shock, the factors are
\begin{align}
\epsilon_\rho &= \frac{1}{\epsilon_v} 
= \frac{(\gamma +1)\mathcal{M}^2}{(\gamma -1)\mathcal{M}^2 + 2}, \label{HR_rho}\\
\epsilon_p &= \frac{2\gamma \mathcal{M}^2 - (\gamma -1)}{\gamma +1}, \label{HR_p}\\
\epsilon_{c_\mathrm{s}} &= \left( \frac{\epsilon_p}{\epsilon_\rho} \right)^{1/2}. \label{HR_cs}
\end{align}
The post-shock gas properties serve as the OBCs for the viscous accretion disc onto the CO when $\mathcal{M}>1$.

\subsection{General formulae for accretion properties}\label{general}

We consider the general case where both the CO and the surrounding disc gas may deviate from Keplerian rotation and possess non-zero radial velocity components relative to the SMBH.
We still assume that the orbits are coplanar; non-coplanar configurations are not considered here (see section~\ref{summary} for further discussion).

Let the CO and the local disc gas orbit the SMBH with angular velocities $f_\mathrm{CO}\Omega_\mathrm{K}$ and $f_\mathrm{gas}\Omega_\mathrm{K}$, respectively, where $\Omega_\mathrm{K} = \sqrt{GM_1/R^3}$. 
The difference in the radial velocity components in the SMBH rest frame is denoted as $v_{R,\mathrm{rel}} = v_{R,\mathrm{gas}} - v_{R,\mathrm{CO}}$.
In the non-inertial frame comoving with the CO, the velocity of the disc gas near the CO can be expressed as
\begin{equation}\label{vgas2}
    \bm{v}_\mathrm{gas}= [f_\mathrm{gas}\Omega_\mathrm{K} - f_\mathrm{CO}S ]R \bm{e}_\Phi + v_{R,\mathrm{rel}} \bm{e}_R,
\end{equation}
where $S \equiv \Omega_\mathrm{K}(R_2)=\sqrt{GM_1/R_2^3}$, a constant value for given $M_1$ and $R_2$, and $f_\mathrm{CO}S$ represents the angular velocity of the CO as well as the comoving frame.
The bulk relative velocity is thus 
\begin{equation}\label{vbulk2}
    \bm v_{\rm bulk} = \Delta f S R_2 \bm{e}_\Phi + v_{R,\mathrm{rel}} \bm{e}_R,
\end{equation}
where $\Delta f \equiv f_\mathrm{gas}-f_\mathrm{CO}$, and $S R_2$ is the Keplerian velocity around the SMBH at radius $R_2$.

The velocity-gradient contribution, $\bm{v}_\mathrm{grad}$, follows by substituting $\bm{v}_\mathrm{gas}$ into equations~\eqref{eq_dv1}--\eqref{eq_dv3}.
Because the CO’s accretion region is much smaller than its orbital radius ($\Delta R \ll R_2$), we treat $f_\mathrm{gas}$ and $v_{R,\mathrm{rel}}$ as locally constant.
Written in the local Cartesian coordinate system defined in section~\ref{accretion}, the components of $\bm{v}_\mathrm{grad}$ are
\begin{align}
v_{x,\mathrm{grad}} &\approx  - A S y + \frac{v_{R,\mathrm{rel}}}{R_2}x, \label{grad_vx2} \\
v_{y,\mathrm{grad}} &\approx - \Delta f S x, \label{grad_vy2} \\
v_{z,\mathrm{grad}} &\approx 0, \label{grad_vz2}
\end{align}
where we define $A \equiv f_\mathrm{gas}/2 + f_\mathrm{CO}$.
For $f_\mathrm{gas}=f_\mathrm{CO}=1$ and $v_{R,\mathrm{rel}}=0$, these reduce to equations~\eqref{coro_v1}--\eqref{coro_v3}.

%---------subsonic case-----------
\subsubsection{Subsonic regime}
In the subsonic regime ($\mathcal{M}<1$), the flow is smoothly focused toward the CO, and the angular velocity associated with $\bm{v}_\mathrm{grad}$ (with counterclockwise rotation defined as positive) is
\begin{equation}\label{omega}
\omega = \frac{x v_{y,\mathrm{grad}} - y v_{x,\mathrm{grad}}}{x^2 + y^2}.
\end{equation}
Using the formalism of Appendix~\ref{app:max_omega}, $\omega$ can be expressed in terms of three coefficients $K$, $M$, and $N$, with the maximum magnitude on a circular orbit given by
\begin{equation}
|\omega|_{\max} = |K| + \sqrt{M^2+N^2}.
\end{equation}
Combined with equations~\eqref{grad_vx2} and \eqref{grad_vy2}, we obtain
\begin{equation}\label{omega_max42}
|\omega|_{\max} = \eta S,
\end{equation}
where
\begin{equation}\label{eta_42}
\eta= \frac{1}{2} \left( |A - \Delta f| + \sqrt{ (A + \Delta f)^2 + \left( \frac{v_{R,\mathrm{rel}}}{S R_2} \right)^2 } \right).
\end{equation}
Equating $|\omega|_{\max}$ to the Keplerian angular velocity then gives the Keplerian radius of the CO disc,
\begin{equation}\label{rk_41}
r_\mathrm{K} = \left( \frac{3}{\eta^2} \right)^{1/3} r_\mathrm{H}.
\end{equation}
As a consistency check, for $f_\mathrm{gas}=f_\mathrm{CO}=1$ and $v_{R,\mathrm{rel}}=0$ we obtain $\eta=3/2$, which leads to $r_\mathrm{K} = (4/3)^{1/3} r_\mathrm{H}$, thereby recovering equation~\eqref{eq_rK} in the circular Keplerian case.

As in section~\ref{accretion}, we calculate the viscous accretion rate from equation~\eqref{eq_mdotH} by setting the outer boundary to the Hill radius ($r_\mathrm{out} = r_\mathrm{H}$) and using the maximum angular velocity $\omega(r_\mathrm{out})=\eta S$. The Keplerian angular velocity at this boundary, $\omega_\mathrm{K}(r_\mathrm{out})=\omega_\mathrm{K}(r_\mathrm{H})=\sqrt{3} S$, is unchanged. 
Applying the derivation from section~\ref{accretion} yields
\begin{equation}\label{eq_mdotvis_eta}
    \dot{M}_\mathrm{vis}
    =\alpha \frac{\sqrt{3}}{\eta} \left(\frac{r_\mathrm{H}}{r_\mathrm{B}}\right)^3\dot{M}_\mathrm{B}.
\end{equation}
As a further check, setting $f_\mathrm{gas}=f_\mathrm{CO}=1$ and $v_{R,\mathrm{rel}}=0$ gives $\eta=3/2$, reducing equation~\eqref{eq_mdotvis_eta} to equation~\eqref{eq_mdotvis_rH}.

%---------supersonic case-----------
\subsubsection{Supersonic regime}\label{supersonic}

In the supersonic regime ($\mathcal{M}>1$), the inclusion of CO gravity leads to the formation of a bow shock, with its surface approximated as perpendicular to $\bm{v}_\mathrm{bulk}$, as discussed above.
Across the shock, the tangential velocity component is unchanged, while the normal component is reduced by the factor $\epsilon_v = [(\gamma -1)\mathcal{M}^2 + 2] / [(\gamma +1)\mathcal{M}^2]$ (see equation~\eqref{HR_rho}). 
To simplify the post-shock analysis, we rotate the coordinate system in the disc plane by an angle $\theta$ about the $z$-axis so that the new $\bar{x}$-axis is aligned with $\bm v_{\rm bulk}$. We define
\begin{equation}\label{defs}
B \equiv \Delta f S R_2, \qquad
V_{\rm b} \equiv \lvert\bm v_{\rm bulk}\rvert =  \sqrt{B^2+v_{R,\mathrm{rel}}^2},
\end{equation}
so that
\begin{equation}\label{cos_sin_theta}
\cos\theta = \frac{B}{V_{\rm b}},\qquad
\sin\theta = \frac{v_{R,\mathrm{rel}}}{V_{\rm b}}.
\end{equation}

In the rotated coordinates $(\bar x,\bar y,\bar z)$, the velocity components of $\bm{v}_\mathrm{grad}$ are
\begin{align}
\bar v_{x,\mathrm{grad}} &\approx
- AS\Big(\cos\theta\sin\theta\,\bar x + \cos^2\theta\,\bar y\Big),
\label{vbarx_barcoords}\\
\bar v_{y,\mathrm{grad}} &\approx
-\frac{V_{\rm b}}{R_2}\Big(\cos\theta\bar x - \sin\theta\bar y\Big)
+ AS\Big(\sin^2\theta\bar x +  \sin\theta\cos\theta\bar y\Big).
\label{vbary_barcoords}
\end{align}
The $z$-component is zero in both the original and the rotated coordinates and is therefore omitted.
The corresponding post-shock components are
\begin{align}
\bar v'_{x,\mathrm{grad}}
&\approx -AS\big(\cos\theta\sin\theta \bar x + \cos^2\theta \bar y\big) \epsilon_v, \label{grad_vx'2} \\
\bar v'_{y,\mathrm{grad}}
&= \bar v_{y,\mathrm{grad}}.
\label{grad_vy'2}
\end{align}

The maximum post-shock angular velocity on a circular orbit can still be written as 
$|\omega|_{\max} = \eta S$, but with $\eta$ updated to incorporate the velocity jump conditions.
Consequently, the formula for the Keplerian radius continues to follow equation~\eqref{rk_41}, now with the updated $\eta$.
Combining the formalism of Appendix~\ref{app:max_omega} with 
equations~\eqref{grad_vx'2} and \eqref{grad_vy'2}, we obtain
\begin{multline}\label{eta'42}
\eta = \frac{1}{2} \biggl\{
\left| - \Delta f + A\left(\sin^2\theta + \epsilon_v\cos^2\theta\right) \right| +\\
\sqrt{
\left[ - \Delta f + A\left(\sin^2\theta - \epsilon_v\cos^2\theta\right) \right]^2
+ \left[ \tfrac{v_{R,\mathrm{rel}}}{S R_2} + A(1+\epsilon_v)\sin\theta\cos\theta \right]^2}
\biggr\}.
\end{multline}
As a consistency check, for $\mathcal{M}=1$ we have $\epsilon_v=1$, and equation~\eqref{eta'42} reduces to \eqref{eta_42}, which follows upon using the identity $v_{R,\mathrm{rel}}/(S R_2)=\Delta f\tan\theta$ from equations~\eqref{defs}--\eqref{cos_sin_theta}.
This confirms that our expression for $\eta$ is continuous across the transition from subsonic to supersonic regimes.

While the expressions for the angular velocities at the outer boundary $r_\mathrm{out}=r_\mathrm{H}$, namely $\omega(r_\mathrm{out})=\eta S$ and $\omega_\mathrm{K}(r_\mathrm{out})=\sqrt{3} S$, retain their previous form, the OBCs for density and sound speed are replaced by the post-shock values $\rho'$ and $c_\mathrm{s}'$. According to equation~\eqref{eq_mdotH}, this substitution introduces an additional factor in the expression for $\dot{M}_\mathrm{vis}$,
\begin{equation}\label{HR_mdot}
\epsilon_{\dot{M}}
= \epsilon_\rho \, \epsilon_{c_\mathrm{s}}^3
= \frac{\bigl[\,2\gamma \mathcal{M}^2 - (\gamma -1)\,\bigr]^{3/2} 
\bigl[\,(\gamma -1)\mathcal{M}^2+2\,\bigr]^{1/2}}
{(\gamma+1)^2 \, \mathcal{M}} .
\end{equation}
The viscous accretion rate thus becomes
\begin{equation}\label{eq_mdotvis_eta'}
    \dot{M}_\mathrm{vis}
    =\alpha \frac{\sqrt{3}}{\eta} \epsilon_{\dot{M}}\left(\frac{r_\mathrm{H}}{r_\mathrm{B}}\right)^3\dot{M}_\mathrm{B}.
\end{equation}

Using equations~\eqref{rBHL_bulk} and \eqref{BHL_rate_bulk}, the same expression can be recast as
\begin{equation}\label{eq_mdotvis_eta'_BHL}
    \dot{M}_\mathrm{vis}
    =\alpha \frac{\sqrt{3}\epsilon_{\dot{M}}}{\eta (1+\mathcal{M}^2)^{3/2}} \left(\frac{r_\mathrm{H}}{r_\mathrm{BHL}}\right)^3\dot{M}_\mathrm{BHL}.
\end{equation}
Equations~\eqref{eq_mdotvis_eta'} and \eqref{eq_mdotvis_eta'_BHL} are mathematically equivalent, but the latter is more convenient for analyzing high-Mach-number regimes. In this form, the factor $\sqrt{3}\epsilon_{\dot{M}}/[\eta (1+\mathcal{M}^2)^{3/2}]$ carries the same leading-order dependence on $\mathcal{M}$ in numerator and denominator, preventing the coefficient from becoming artificially large when $\mathcal{M}$ is high.

\subsubsection{Unified formula for the viscous accretion rate}\label{uni_form}

The viscous accretion rate can be written in a unified form as
\begin{equation}\label{mdotvis_uni}
\dot{M}_\mathrm{vis}
    = \alpha \, \xi \left(\frac{r_\mathrm{H}}{r_\mathrm{BHL}}\right)^3 \dot{M}_\mathrm{BHL},
\end{equation}
where $\xi$ is a dimensionless coefficient. 
We adopt the BHL radius and accretion rate, rather than the Bondi values, to ensure that $\xi$ remains bounded for $\mathcal{M}\gg1$.\footnote{
In this form, $\xi$ stays of order unity even at high Mach numbers (as shown in section~\ref{applications}), and can therefore be set to unity in practical, order-of-magnitude estimates.}

Since $r_\mathrm{H}$, $r_\mathrm{BHL}$, and $\dot{M}_\mathrm{BHL}$ are all well-defined quantities 
(see equations~\eqref{rH}, \eqref{rBHL_bulk}, and \eqref{BHL_rate_bulk}), 
the problem reduces to specifying the form of $\xi$, which is
\begin{equation}\label{xiH}
\xi = 
\begin{cases}
\dfrac{\sqrt{3}}{\eta(1+\mathcal{M}^2)^{3/2}}, & \mathcal{M} < 1, \\[10pt]
\dfrac{\sqrt{3}\epsilon_{\dot{M}}}{\eta(1+\mathcal{M}^2)^{3/2}}, & \mathcal{M} \geq 1 ,
\end{cases}
\end{equation}
where $\eta$ is given by equations~\eqref{eta_42} and \eqref{eta'42} in the subsonic and supersonic regimes, respectively, and is continuous at $\mathcal{M}=1$ as discussed in section~\ref{supersonic}.
The shock jump factor $\epsilon_{\dot{M}}$ is given by equation~\eqref{HR_mdot} and approaches unity at $\mathcal{M}=1$. 
Therefore, $\xi$ -- and consequently $\dot{M}_\mathrm{vis}$ -- varies smoothly across the subsonic and supersonic regimes.
As a consistency check, for the circular Keplerian case in section~\ref{accretion} with no bulk relative motion ($\mathcal{M}=0$, $\eta=3/2$, $r_\mathrm{BHL}=r_\mathrm{B}$, and $\dot{M}_\mathrm{BHL}=\dot{M}_\mathrm{B}$), we obtain $\xi=2/\sqrt{3}$, recovering equation~\eqref{eq_mdotvis_rH}.

%---------------
We can derive a direct relation between $\dot{M}_\mathrm{vis}$ and $\dot{M}_1$ in the general case, similar to equation~\eqref{mdot1_mdotvis} in section~\ref{accretion}.
The accretion rate of the AGN disc is now
\begin{equation}\label{mdot1_general}
    \dot{M}_1 = \frac{4\alpha \pi \rho_\mathrm{d}c_\mathrm{sd}^3}{f_\mathrm{gas} S^2},
\end{equation}
valid for $R_2 \gg R_\mathrm{in} \sim 3R_\mathrm{g}$.
The viscous accretion rate onto the CO can be expressed as
\begin{equation}
\dot{M}_\mathrm{vis} =
\dfrac{4\alpha \pi \rho_\mathrm{d} c_\mathrm{sd}^3}{\sqrt{3}\,\eta S^2}
\times
\begin{cases}
1, & \mathcal{M} < 1, \\[6pt]
\epsilon_{\dot{M}}, & \mathcal{M} \geq 1,
\end{cases}
\end{equation}
which leads to
\begin{equation}\label{mdot1_mdotvis_general}
\dot{M}_\mathrm{vis} =
\dfrac{f_\mathrm{gas}}{\sqrt{3} \eta} \dot{M}_1
\times
\begin{cases}
1, & \mathcal{M} < 1, \\[6pt]
\epsilon_{\dot{M}}, & \mathcal{M} \geq 1.
\end{cases}
\end{equation}
This expression reduces to equation~\eqref{mdot1_mdotvis} in the circular Keplerian case ($\mathcal{M}=0$, $f_\mathrm{gas}=1$, and $\eta=3/2$).
%

%---------------
In practice, the accretion rate onto the CO should be taken as the minimum of the viscous and BHL rates:
\begin{equation}\label{mdot_final}
    \dot{M}_\mathrm{CO} = \min\{\dot{M}_\mathrm{vis}, \dot{M}_\mathrm{BHL}\}.
\end{equation}
It is viscosity-limited for $\dot{M}_\mathrm{vis} < \dot{M}_\mathrm{BHL}$, which sets a lower limit on the mass ratio $q \equiv M_2/M_1$, as demonstrated below.

Note that the CO accretion disc, like the Bondi and BHL accretion flows, is vertically confined by the local AGN disc scale height, which imposes an additional constraint on the accretion rate.
This effectively corresponds to a change in the shock-induced factor $\epsilon_{\dot{M}}$.
A detailed discussion of this geometric limit is provided in section~\ref{summary}.

\subsubsection{Lower limit on mass ratio for viscous accretion}\label{lower_q}

The CO accretion is viscosity-limited for $\dot{M}_\mathrm{vis} < \dot{M}_\mathrm{BHL}$, which, according to equation~\eqref{mdotvis_uni}, implies
\begin{equation}\label{q1}
\alpha \xi \left(\frac{r_\mathrm{H}}{r_\mathrm{BHL}}\right)^3 <1.
\end{equation}
Substituting equations~\eqref{rH} and \eqref{rBHL_bulk} into the above expression, we obtain
\begin{equation}\label{q2}
q > \left[\frac{\alpha \xi(1+\mathcal{M}^2)^{3}}{3}\right]^{1/2}\left(\frac{c_{\mathrm{sd}}}{V_{\mathrm{K}}}\right)^3,
\end{equation}
where $V_{\mathrm{K}} = \Omega_\mathrm{K} R_2$ is the Keplerian velocity around the SMBH.
Adopting $H = c_\mathrm{sd}/\Omega_\mathrm{K}$, we eventually obtain
\begin{equation}
q > \left[\frac{\alpha \xi(1+\mathcal{M}^2)^{3}}{3}\right]^{1/2} h^3,
\end{equation}
where $h \equiv H/R_2$ is the local aspect ratio of the AGN disc. In typical subsonic cases, the right-hand side can be approximated as $\sqrt{\alpha/3} h^3$. 
Thus, for COs on nearly Keplerian prograde orbits -- for which the bulk relative motion is typically subsonic, as analysed and illustrated in section~\ref{slimdisc} -- the aspect ratio sets the lower limit of $q$ above which the viscous accretion rate applies.

For thin AGN discs such as the SSD model, the aspect ratio is typically $h \sim 10^{-3}$--$10^{-2}$, while $\alpha$ is typically of order $0.1$, yielding a lower limit of $q \sim 10^{-7}$ or lower. This limit is consistent with Fig.~\ref{fig3}, where $q = 10^{-7}$ leads to viscosity-limited accretion except in the inner, hotter regions with larger $h$.

For slim discs, $h \sim 0.1$ or higher depending on the accretion rate, giving a lower limit of $q \sim 10^{-4}$. In this case, accretion onto stellar-mass COs is generally BHL-limited (see section~\ref{slimdisc} for numerical validation).

For ADAFs, the aspect ratio is even higher, typically approaching $h \sim 1$ \citep[e.g.,][]{Yuan2014}, and the accretion onto embedded COs should be determined by the BHL rate. 

%==================================
\subsubsection{Gravitational stability: Toomre $Q$ parameter}
\label{Q_415}
We now generalise the calculation of the Toomre $Q$ parameter presented in section~\ref{accretion}.
Using equation~\eqref{Q_general}, the $Q$ parameter for the AGN disc at the location of the CO is
\begin{equation}\label{Q1_general}
Q_1 =  \frac{f_\mathrm{gas}S^2}{2\pi G \rho_\mathrm{d}}.
\end{equation}

For the viscous accretion disc around the CO, the minimum $Q$ occurs at the outer boundary $r_\mathrm{out}$:
\begin{equation}\label{Q2_general}
Q_2 =
\begin{cases}
\dfrac{\omega(r_\mathrm{out}) \cdot \omega_\mathrm{K}(r_\mathrm{out})}{2\pi G \rho_\mathrm{d}}
= \dfrac{\sqrt{3}\eta S^2}{2\pi G \rho_\mathrm{d}}
=\dfrac{\sqrt{3}\eta}{f_\mathrm{gas}}Q_1, & \mathcal{M} < 1, \\[10pt]
\dfrac{\omega(r_\mathrm{out}) \cdot \omega_\mathrm{K}(r_\mathrm{out})}{2\pi G \rho'}
= \dfrac{\sqrt{3}\eta S^2}{2\pi G \rho_\mathrm{d}\epsilon_\rho}
=\dfrac{\sqrt{3}\eta}{f_\mathrm{gas}\epsilon_\rho}Q_1,  & \mathcal{M} \geq 1.
\end{cases}
\end{equation}
As a consistency check, in the circular Keplerian case discussed in section~\ref{accretion} ($\eta=3/2$ and $f_\mathrm{gas}=1$), we obtain $Q_2 = 3\sqrt{3}/2Q_1$, recovering equation~\eqref{Q2_c3}.

In the general case, $Q_2$ can be smaller than $Q_1$, particularly in the supersonic regime where $\epsilon_\rho > 1$ (see section~\ref{eccen}). 
When the geometric constraint imposed by the AGN disc scale height is taken into account, $Q_2$ may be larger than predicted by equation~\eqref{Q2_general} under certain conditions (see section~\ref{summary}).

%==============================
\subsection{Illustrative examples}\label{applications}
\subsubsection{Slim-disc active galactic nucleus case}\label{slimdisc}

We first consider the case where the AGN disc is described by a slim disc.
We adopt $M_1=10^8M_\odot$, $M_2=10M_\odot$, $\alpha=0.1$, $\dot{M}_1=100L_\mathrm{Edd1}/c^2$, $\mu=0.6$, and $\kappa_\mathrm{es}=0.35$ cm$^2$ g$^{-1}$.
The disc structure is obtained by solving the standard slim-disc equations \citep[e.g.][]{Ab1988,Kato2008}.
The CO is assumed to follow a prograde, circular, Keplerian orbit in the disc mid-plane ($f_\mathrm{CO}=1$, $v_{R,\mathrm{CO}}=0$).
With both the disc structure and the CO orbital parameters specified, we substitute these into our formulae to obtain the corresponding accretion properties.

\begin{figure*}
    \centering
    \includegraphics[width=\textwidth]{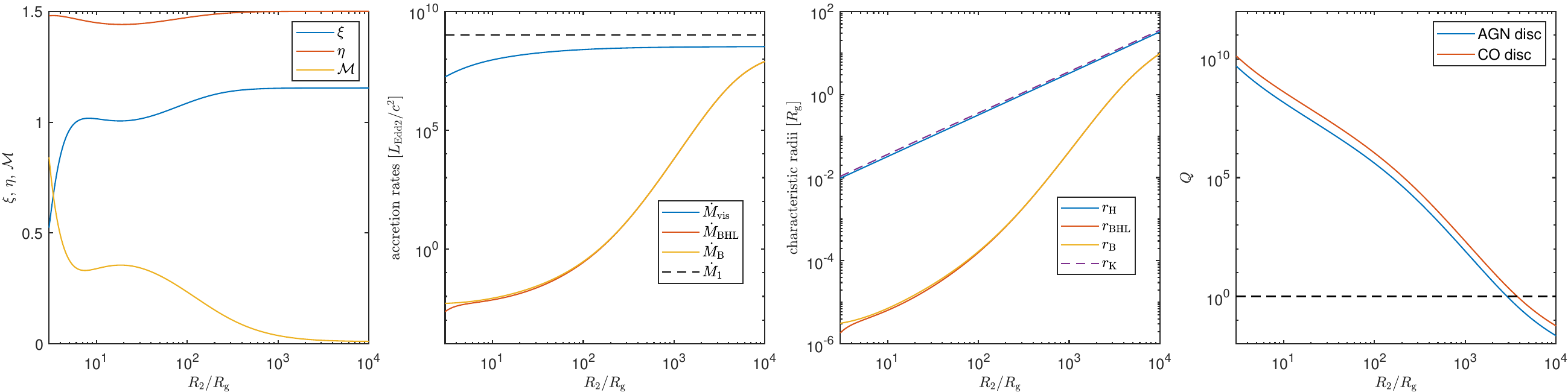}
    \caption{
        Profiles of dimensionless parameters ($\xi$, $\eta$, $\mathcal{M}$), 
        accretion rates ($\dot{M}_\mathrm{vis}$, $\dot{M}_\mathrm{BHL}$, $\dot{M}_\mathrm{B}$), 
        characteristic radii ($r_\mathrm{H}$, $r_\mathrm{BHL}$, $r_\mathrm{B}$, $r_\mathrm{K}$),
        and Toomre $Q$ as functions of the SMBH--CO separation $R_2$,
        for the slim-disc case with $M_1=10^8M_\odot$, $M_2=10M_\odot$, $\alpha=0.1$, 
        $\dot{M}_1=100L_\mathrm{Edd1}/c^2$ (corresponding to $10^9L_\mathrm{Edd2}/c^2$), 
        $\mu=0.6$, and $\kappa_\mathrm{es}=0.35$ cm$^2$ g$^{-1}$.
        The CO is assumed to be on a circular Keplerian orbit ($f_\mathrm{CO}=1$, $v_{R,\mathrm{CO}}\approx0$).
    }
    \label{fig4}
\end{figure*}

Figure~\ref{fig4} presents the results in four panels, showing (from left to right) the dimensionless parameters ($\xi$, $\eta$, and $\mathcal{M}$), the accretion rates ($\dot{M}_\mathrm{vis}$, $\dot{M}_\mathrm{BHL}$, and $\dot{M}_\mathrm{B}$), the characteristic radii ($r_\mathrm{H}$, $r_\mathrm{BHL}$, $r_\mathrm{B}$, and $r_\mathrm{K}$), and the Toomre $Q$ parameters of the AGN and CO discs, all as functions of the SMBH--CO separation $R_2$.
It is evident that the bulk relative motion between the CO and the disc gas remains subsonic at all radii.
This arises mainly because:  
(1) the radial motion of the disc gas remains subsonic except inside the sonic radius, which typically lies near the ISCO, within which the CO can no longer maintain a stable bound orbit; 
(2) although the rotation of the slim disc is sub-Keplerian, producing an azimuthal component of $v_\mathrm{bulk}$, the high disc temperature raises the sound speed and keeps the Mach number below unity.

In this subsonic regime, the influence of bulk relative motion on the viscous accretion rate is relatively small.
Compared with the SSD case shown in Fig.~\ref{fig3}, the increase of $\dot{M}_\mathrm{vis}$ is mainly due to the changes in disc gas properties, namely the density $\rho_\mathrm{d}$ and sound speed $c_\mathrm{sd}$.
Especially in the outer slim disc, $\mathcal{M}$ approaches zero and the system effectively reduces to the scenario discussed in section~\ref{accretion}, with
$\eta \approx 3/2$, $\xi\approx 2/\sqrt{3}$, $\dot{M}_\mathrm{vis} \approx 2/(3\sqrt{3})\,\dot{M}_1$, and $Q_2 \approx 3\sqrt{3}/2 Q_1$.

By contrast, the Bondi and BHL accretion rates are strongly suppressed compared with the SSD case, owing to the much higher slim-disc temperatures caused by photon trapping at high optical depth.
Since both $\dot{M}_\mathrm{B}$ and $\dot{M}_\mathrm{BHL}$ scale as $c_\mathrm{sd}^{-3}$, they are significantly smaller than the viscous accretion rate, particularly in the inner, hotter regions of the disc. 
As a result, accretion onto the CO is limited by the BHL rate. 
This agrees with the prediction in section~\ref{lower_q} that CO accretion in a slim-disc AGN is BHL-limited for $q \lesssim 10^{-4}$.
%-------------------
\subsubsection{Eccentric-orbit compact object case}\label{eccen}
We now consider a CO moving on an eccentric orbit with eccentricity $e$ and semi-major axis $a$ around the SMBH.
The AGN disc is assumed to be an SSD, with Keplerian rotation ($f_\mathrm{gas}=1$) and negligible radial velocity ($v_{R,\mathrm{gas}} \approx 0$).
Let $\nu$ be the true anomaly of the CO measured from the pericentre. The azimuthal and radial components of its velocity can be expressed as
\begin{align}
v_{\Phi,\mathrm{CO}} &= \sqrt{\frac{GM_1}{a}} \frac{1 + e \cos \nu}{\sqrt{1 - e^2}}, \\
v_{R,\mathrm{CO}} &= \sqrt{\frac{GM_1}{a}} \frac{e \sin \nu}{\sqrt{1 - e^2}},
\end{align}
and the distance from the SMBH is given by the orbit equation
\begin{equation}
R_2 = \frac{a(1 - e^2)}{1 + e \cos \nu}.
\end{equation}
The Keplerian velocity around the SMBH at $R_2$ is $S R_2 = \sqrt{GM_1/R_2}$. Substituting the expression for $R_2$ yields
\begin{equation}
S R_2 = \sqrt{\frac{GM_1}{a}} \frac{\sqrt{1 + e \cos \nu}}{\sqrt{1 - e^2}}.
\end{equation}
Consequently we get
\begin{align}
f_\mathrm{CO} &= \frac{v_{\Phi,\mathrm{CO}}}{S R_2}
= \sqrt{1 + e \cos \nu}, \\
\frac{v_{R,\mathrm{rel}}}{S R_2} &= -\frac{v_{R,\mathrm{CO}}}{S R_2} = -\frac{e \sin \nu}{\sqrt{1 + e \cos \nu}}.
\end{align}
Combined with the disc gas properties, we can then obtain the viscous accretion rate.

\begin{figure*}
    \centering
    \includegraphics[width=\textwidth]{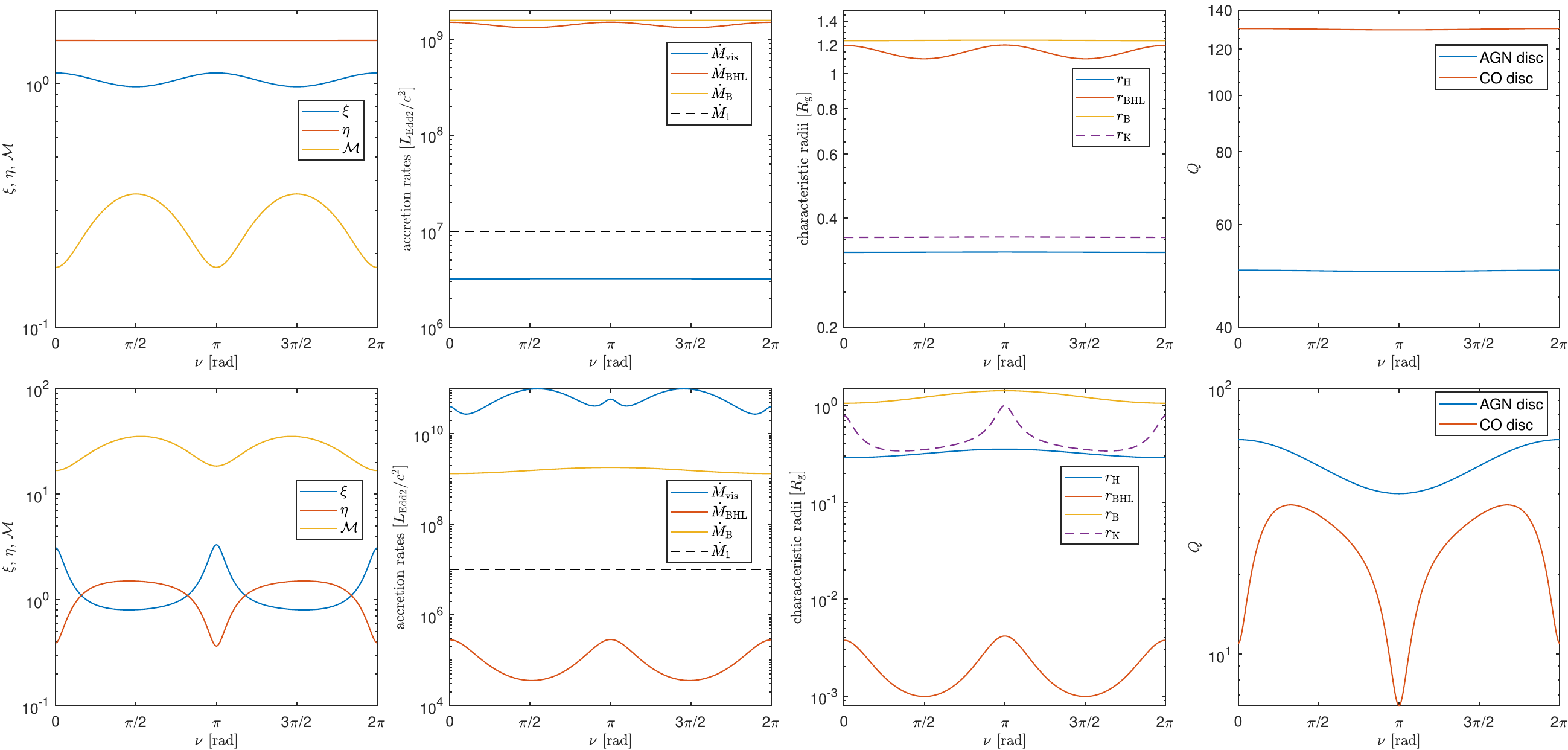}
    \caption{
        Variations of dimensionless parameters ($\xi$, $\eta$, $\mathcal{M}$), 
        accretion rates ($\dot{M}_\mathrm{vis}$, $\dot{M}_\mathrm{BHL}$, $\dot{M}_\mathrm{B}$), 
        characteristic radii ($r_\mathrm{H}$, $r_\mathrm{BHL}$, $r_\mathrm{B}$, $r_\mathrm{K}$),
        and Toomre $Q$ along the CO orbital phase.
        The horizontal axis shows the true anomaly $\nu$, with $\nu=0$ (and $2\pi$) corresponding 
    to pericentre, and $\nu=\pi$ corresponding to apocentre.
        The upper and lower panels correspond to eccentricity $e=0.001$ and $e=0.1$, respectively,
        both with semi-major axis $a=100R_\mathrm{g}$.
        The AGN disc is the same as that in Fig.~\ref{fig3} in section~\ref{accretion}.
    }
    \label{eccentric}
\end{figure*}

Figure~\ref{eccentric} displays the profiles of various accretion properties as functions of $\nu$. The AGN disc is taken to be the same as that in Fig.~\ref{fig3} in section~\ref{accretion}. 
The upper and lower panels correspond to eccentricity $e=0.001$ and $0.1$, respectively, both with $a=100R_\mathrm{g}$.

When the CO orbit is only slightly eccentric ($e=0.001$), the bulk relative motion is subsonic.
For such a low eccentricity, $R_2$ is almost constant at different orbital phases, leading to nearly constant values of $\dot{M}_\mathrm{B}$, $r_\mathrm{B}$, and $r_\mathrm{H}$. The viscous accretion rate $\dot{M}_\mathrm{vis}$ consequently depends solely on $\eta$ (see equation~\eqref{eq_mdotvis_eta}; or, with the unified formula equation~\eqref{mdotvis_uni}, the dependence on the Mach number $\mathcal{M}$ in $r_\mathrm{BHL}$, $\dot{M}_\mathrm{BHL}$, and $\xi$ cancels out), which is close to its value with no bulk relative motion, 3/2, as shown in the upper panels. This leads to a nearly constant value of $\dot{M}_\mathrm{vis}$, which is almost the same as that in Fig.~\ref{fig3} for $R_2=a$.
With $\eta \approx 3/2$, we also obtain
$Q_2\approx 3\sqrt{3}/2Q_1$, such that the CO disc is self-gravitationally stable provided that the AGN disc at that location is stable.
Thus, for very low eccentricities, the accretion behaviour converges to the circular-orbit limit, as expected.

On the other hand, for $e=0.1$, the bulk relative motion becomes significantly supersonic.
Consequently, the BHL radius $r_\mathrm{BHL}$ becomes much smaller than the Hill radius $r_\mathrm{H}$, and the accretion onto the CO is governed by the BHL rate, as seen in the lower panels.
The viscous accretion rate in this case becomes much larger than both the Bondi and BHL rates.
This behaviour arises because both density and sound speed of the disc gas increase significantly across the shock due to the high Mach numbers, such that the shock-induced factor $\epsilon_{\dot{M}}$ becomes very large.
Even if the geometric constraint imposed by the AGN disc scale height is considered (as detailed in section~\ref{summary}), it only mitigates the increase in $\dot{M}_\mathrm{vis}$, and thus the conclusion of a BHL-limited accretion remains unchanged.
We note that $\xi$ remains of order unity, since we incorporated another $(1+\mathcal{M}^2)^{-3/2}$ factor into $\xi$ by adopting $r_\mathrm{BHL}$ and $\dot{M}_\mathrm{BHL}$ rather than $r_\mathrm{B}$ and $\dot{M}_\mathrm{B}$ in the unified formula (see sections~\ref{uni_form}).
The parameter $Q_2$ becomes less than $Q_1$ due to the increased density post-shock (see equation~\eqref{Q2_general}). However, in this highly supersonic regime, the accretion onto the CO is dominated by the BHL mechanism rather than by the viscous process, and $Q_2$ is no longer relevant.

%%%%%%%%%%%%%%%%%%%%%%%%%%%%%%%%%%%%%%%%%%%%%%%%%%%%%%%
\section{Summary and discussion}\label{summary}

We present a new framework that provides a more physically motivated estimate of the accretion rate onto a CO embedded in an AGN disc.
The model self-consistently accounts for the gas--CO relative motion by decomposing it into a local gradient term (due to differential rotation) and bulk motion (from differing orbital parameters). The former carries net angular momentum, leading to the formation of a viscous accretion disc around the CO whose rate is given by $\dot{M}_\mathrm{vis} = \alpha \xi (r_\mathrm{H}/r_\mathrm{BHL})^3\dot{M}_\mathrm{BHL}$, with $\xi$ depending on the specific dynamical configuration. 
Alternatively, the accretion rate can be linked to global AGN properties via $\dot{M}_\mathrm{vis} = f_\mathrm{gas}\epsilon_{\dot{M}}/(\sqrt{3} \eta) \dot{M}_1$ (valid for $R_2 \gg R_\mathrm{in} \sim 3R_\mathrm{g}$), where the shock-induced factor $\epsilon_{\dot{M}}$ takes unity in the subsonic regime.
In the circular Keplerian case presented in section~\ref{accretion}, this yields $\dot{M}_\mathrm{vis} \approx 0.38 \dot{M}_1$.
The actual accretion rate should be taken as $\dot{M}_\mathrm{CO} = \min\{\dot{M}_\mathrm{vis}, \dot{M}_\mathrm{BHL}\}$. It is viscosity-limited rather than BHL-limited for $q > [\alpha \xi(1+\mathcal{M}^2)^{3}/3]^{1/2} h^3$.
Regarding self-gravitational stability, the minimum Toomre $Q$ of the CO disc typically exceeds that of the AGN disc at the location of the CO for subsonic bulk relative motion, though for supersonic cases the stability should be evaluated explicitly with equation~\eqref{Q2_general}.

As the mass ratio $q$ increases, the CO's tidal torque can overcome the disc's viscous torque, leading to gap opening in the AGN disc and thereby quenching the gas supply.
A commonly used critical mass ratio for gap opening is
\begin{equation}
    q_\mathrm{c} \sim \max\{ 3h^{3},\, C\alpha h^{2} \},
\end{equation}
where $C\sim25$--$50$ \citep{Lin1986,Bryden1999,Crida2006}. 
For typical thin AGN discs ($h\sim10^{-3}$--$10^{-2}$, $\alpha\sim0.1$), this corresponds to $q_\mathrm{c} \sim 10^{-4}$, and gap opening happens for $q \gtrsim q_\mathrm{c}$, consistent with previous studies (e.g., \citealt{McK2014}). 
Consequently, the validity of our model is limited to systems with \(q \lesssim q_\mathrm{c}\), where the gas supply remains uninterrupted.

The classical Bondi and BHL rates are suppressed when their characteristic radii exceed the local half-thickness of the AGN disc, $H_1 = c_\mathrm{sd}/S$. 
In our model, the CO accretion disc has a maximum half-thickness at its outer boundary $r_\mathrm{H}$, given by $H_{2,\mathrm{max}} = c_\mathrm{sd} / \omega_\mathrm{K}(r_\mathrm{H})$ for $\mathcal{M} < 1$ and $H_{2,\mathrm{max}} = \epsilon_{c_\mathrm{s}} c_\mathrm{sd} / \omega_\mathrm{K}(r_\mathrm{H})$ for $\mathcal{M} > 1$. 
Given that $\omega_\mathrm{K}(r_\mathrm{H}) = \sqrt{3}S$, the CO disc is always thinner than the AGN disc ($H_{2,\mathrm{max}} = H_1/\sqrt{3}$) in the subsonic regime. 
This changes in the supersonic regime, where the elevated post-shock sound speed puffs up the CO disc, allowing it to surpass $H_1$ when $\epsilon_{c_\mathrm{s}} > \sqrt{3}$.
Consequently, in this regime, $\dot{M}_\mathrm{vis}$ must be reduced by a factor of $\sqrt{3}/\epsilon_{c_\mathrm{s}}$ to account for the geometric constraint imposed by the AGN disc, while the Toomre parameter $Q_2$ increases by a factor of $\epsilon_{c_\mathrm{s}}/\sqrt{3}$ due to the constrained thickness of the CO disc.
The reduction in $\dot{M}_\mathrm{vis}$ effectively modifies the shock-induced factor $\epsilon_{\dot{M}}$ as follows:
\begin{equation}\label{HR_mdot_eff}
\epsilon_{\dot{M},\mathrm{eff}} = \epsilon_\rho \epsilon_{c_\mathrm{s}}^3 \cdot \frac{\sqrt{3}}{\epsilon_{c_\mathrm{s}}} = \sqrt{3}\,\epsilon_\rho  \epsilon_{c_\mathrm{s}}^2.
\end{equation}
This modification only mitigates -- rather than eliminates -- the shock-induced increase in $\dot{M}_\mathrm{vis}$.
For a typical $\gamma=5/3$, equation~\eqref{HR_cs} shows that this constraint becomes relevant when $\mathcal{M} > 2.62$.

In real astrophysical systems, the final accretion rate onto the CO may be further reduced by radiation feedback and outflows. Radiation force limits the emergent luminosity of spherical accretion to the Eddington value $L_\mathrm{Edd}$, thereby capping the accretion rate at  $L_\mathrm{Edd}/(\eta c^2)$. However, disc accretion permits super-Eddington luminosities \citep{Kato2008,Yuya2016}. Concurrently, photon trapping significantly reduces $\eta$, rendering hyper-Eddington accretion rates ($\gtrsim 5000L_\mathrm{Edd2}/c^2$) plausible even under luminosity constraints \citep{Kohei2016,Yuya2016}. For super-Eddington accretion, the slim disc model predicts an emergent luminosity of \citep{Wang2014}
\begin{equation}\label{L_slim}
L \approx 2 L_{\mathrm{Edd}}[1+\ln (\dot{m} / 50)],
\end{equation}
where $\dot{m}$ is the accretion rate in units of $L_\mathrm{Edd}/c^2$. 
This luminosity saturates at high $\dot{m}$, making it impractical to define a robust upper limit for the accretion rate based solely on emergent radiation.
Instead, because of anisotropic emission in an accretion disc, radiation likely constrains accretion primarily through outflows, driven by the strong radiation forces at high accretion rates, as predicted by both theoretical works \citep[e.g.,][]{Gu2007,HD07,Jiao09,JW11,Begelman2012,Feng2019,CG2022} and numerical simulations \citep[e.g.,][]{Ohsuga2005, Ohsuga2009, OM2007, OM2011, Yang2014, Sadowski2014, Sadowski2015, McKinney2014, Jiang2014, Jiang2019, Jiao15, Kitaki2018, Kitaki2021, Asahina2022,Hu2022}.
The outflow strength remains debated and is commonly parameterised as $\dot{M} \propto r^{s}$. Theoretical work by \cite{Begelman2012} proposed $s \lesssim 1$ for accretion flows with small radiative losses. Numerical simulations of super-Eddington accretion flows typically have $s \sim 0.5-1$, though exceptions exist (e.g., $s=1.7$ in \citealt{McKinney2014}; see \citealt{Jiao2023} for a review of $s$ values). In previous studies, \cite{Pan2021} adopted $s=1$, \cite{Zhang2024} adopted $s=0.5$, and \cite{CRD2023} considered three values of $s=0.2$, 0.5, and 0.8.
Within this simplified framework, the effective accretion rate onto the CO in our model can be estimated as $\dot{M}_\mathrm{eff}=\dot{M}_\mathrm{vis}(r_0/r_\mathrm{out})^s$, where $r_0$ represents the radius inside which outflows become negligible. According to numerical simulations, $r_0 \sim 10r_\mathrm{g}$ ($r_\mathrm{g}=2GM_2/c^2$ is the gravitational radius of the CO).
More precise constraints will require dedicated numerical simulations implementing the initial conditions for the gas--CO relative motion, as presented in sections~\ref{rel_motion} and~\ref{newsec}.

In section~\ref{slimdisc}, we argued that applying the slim disc model to the AGN disc does not introduce supersonic bulk relative motion for a CO on a nearly Keplerian prograde orbit. The same reasoning holds for the ADAF model, which also features a sonic radius near the ISCO and very high temperatures (even higher than in the slim-disc case). Since other widely adopted AGN disc models (see section~\ref{intro}) are Keplerian, the bulk relative motion is likewise expected to remain subsonic under such orbital configurations. 
Given that in this subsonic regime $\dot{M}_\mathrm{vis} = f_\mathrm{gas} /(\sqrt{3} \eta) \dot{M}_1 \sim 0.38 \dot{M}_1$ (or lower in the inner AGN disc; see Figs.~\ref{fig3} and \ref{fig4}), the influence of CO accretion on the AGN disc can typically be neglected in rough estimates.
However, this approximation breaks down if numerous stellar-mass COs are embedded in the AGN disc, as their summed accretion rates could surpass $\dot{M}_1$. In such cases, radiation feedback and outflows become essential considerations, which could greatly suppress effective accretion rates onto COs as discussed above. 
Concurrently, these feedback effects could profoundly alter the structure of the AGN disc. For example, \cite{Gil2022} and \cite{EM2025} assumed that many stellar-mass BHs are embedded in the AGN disc, and found that radiation generated by their accretion can provide enough heating to maintain the self-gravitational stability of the AGN disc in its outer regions, in place of the heating traditionally attributed to star formation \citep{Sirko2003}. A coupled study of the AGN and the CO accretion discs will be the subject of future work.

Our analysis assumes that the CO orbits are coplanar with the AGN disc. This is a well-justified assumption for embedded COs, as recent studies have demonstrated that inclined orbits undergo rapid hydrodynamic damping into the disc plane on short timescales \citep[e.g.,][]{Dit2024,Rowan2025,WH2025}. While our formalism provides a foundation for analyzing the local velocity field in more complex scenarios, a detailed investigation of accretion onto COs on initially inclined orbits -- which involves additional complexities such as vertical shear and out-of-plane forces -- is beyond the scope of this study.

Our model provides a physically motivated description of accretion onto COs embedded in AGN discs, offering critical input for several aspects of the AGN disc channel discussed in section~\ref{intro}.
Unlike the Bondi or BHL prescriptions, it self-consistently incorporates the angular momentum imparted by differential rotation in the AGN disc.
Furthermore, while several previous studies capped the accretion rate at the Eddington value, our framework naturally allows for the inclusion of outflow corrections from the literature, thereby enabling a more realistic treatment of super-Eddington flows.
In addition, it specifies its applicable range of mass ratios and provides a convenient scaling relation between the CO and global AGN accretion rates, making it broadly applicable in population-level studies.
Taken together, these advances establish a physical basis for assessing CO evolution and associated observables within the AGN disc channel, effectively bridging theory with potential observations.

We outline below the direct implications of our findings for the AGN disc channel.
One important consequence is that, by providing a more reliable accretion rate, our model enables a quantitative reassessment of the growth of stellar-mass COs into the pair-instability mass gap, thereby strengthening predictions for the mass spectrum of hierarchical mergers detectable by GW observatories \citep[e.g.,][]{Yang2019, Frag2022}.
Another implication is that the revised accretion-driven mass evolution directly modifies the migration timescales and gas-assisted binary formation processes of COs, offering a more solid foundation for models of binary formation and dynamics in AGN discs \citep[e.g.,][]{Stone2017, LiR2023, DeL2023}, which in turn are critical inputs for future population synthesis studies.
Our framework also places a physically motivated upper limit on the accretion luminosity, thereby constraining the expected brightness and energetics of EM counterparts such as QPEs, X-ray flares, and post-merger afterglows, which is crucial for assessing their detectability and for distinguishing the AGN channel from other formation scenarios \citep[e.g.,][]{Zhu2021, Kimura2021, Perna2021, Linial2023}.
Furthermore, by quantifying the gas consumption and potential feedback from embedded COs, our framework informs studies of AGN disc structure, chemical enrichment, and episodic activity \citep[e.g.,][]{Cant2021,Gil2022,WM2023,ZS2024}.  
By addressing these diverse phenomena, our work strengthens the physical basis of the AGN disc channel, providing a framework to link CO accretion physics with GW sources, EM signatures, and AGN evolution in a self-consistent manner.

\section*{Acknowledgements}

We thank the anonymous reviewer for their constructive comments and valuable suggestions that helped improve the manuscript.
This work was supported by the Science Foundation of Yunnan Province (Nos. 202401AS070046 and 202503AP140013), the International Partnership Program of the Chinese Academy of Sciences (No. 020GJHZ2023030GC), and the Yunnan Revitalization Talent Support Program.

%%%%%%%%%%%%%%%%%%%%%%%%%%%%%%%%%%%%%%%%%%%%%%%%%%
\section*{Data Availability}

The data underlying this article will be shared on reasonable request
to the corresponding author.



\bibliographystyle{mnras}
\bibliography{refs} % if your bibtex file is called example.bib
%%%%%%%%%%%%%%%%% APPENDICES %%%%%%%%%%%%%%%%%%%%%

\appendix
\section{Maximum angular velocity in a linear velocity field}\label{app:max_omega}

Consider a linear velocity field in a Cartesian coordinate system \((x, y)\):
\begin{align}
v_x &= c_{x1} \, x + c_{x2} \, y, \\
v_y &= c_{y1} \, x + c_{y2} \, y.
\end{align}

The angular velocity about the origin is defined as
\begin{equation}
\omega = \frac{x v_y - y v_x}{x^2 + y^2}.
\end{equation}

Substituting the velocity components, we have
\begin{equation}
\omega = \frac{c_{y1} x^2 + (c_{y2} - c_{x1}) xy - c_{x2} y^2}{x^2 + y^2}.
\end{equation}

On a circular path of radius $r$ ($r^2=x^2+y^2$ centred at the origin, we get
\begin{equation}
\omega = \frac{c_{y1} x^2 + (c_{y2} - c_{x1}) xy - c_{x2} y^2}{r^2}.
\end{equation}

Parameterising the circle using \(x = r \cos \phi\) and \(y = r \sin \phi\), we find
\begin{equation}
\omega(\phi) = c_{y1} \cos^2 \phi + (c_{y2} - c_{x1}) \cos \phi \sin \phi - c_{x2} \sin^2 \phi.
\end{equation}

Using trigonometric identities,
\begin{equation}
\begin{aligned}
\cos^2 \phi &= \frac{1 + \cos 2\phi}{2}, \\
\sin^2 \phi &= \frac{1 - \cos 2\phi}{2}, \\
\cos \phi \sin \phi &= \frac{\sin 2\phi}{2}.
\end{aligned}
\end{equation}

we can rewrite
\begin{equation}
\omega(\phi) = \frac{c_{y1} - c_{x2}}{2} + \frac{c_{y1} + c_{x2}}{2} \cos 2\phi + \frac{c_{y2} - c_{x1}}{2} \sin 2\phi.
\end{equation}

Defining
\begin{equation}\label{kmn}
K = \frac{c_{y1} - c_{x2}}{2}, \quad
M = \frac{c_{y1} + c_{x2}}{2}, \quad
N = \frac{c_{y2} - c_{x1}}{2},
\end{equation}
we have
\begin{equation}
\omega(\phi) = K + M \cos 2\phi + N \sin 2\phi.
\end{equation}

The maximum angular velocity magnitude is then
\begin{equation}
|\omega|_\mathrm{max} = |K| + \sqrt{M^2 + N^2},
\end{equation}
and the corresponding angle \(\phi\) is obtained by solving
\begin{equation}
\omega(\phi) = \pm \big(|K| + \sqrt{M^2 + N^2}\big).
\end{equation}

%%%%%%%%%%%%%%%%%%%%%%%%%%%%%%%%%%%%%%%%%%%%%%%%%%

% Don't change these lines
\bsp	% typesetting comment
\label{lastpage}
\end{document}